\documentclass[11pt,epsf,curves]{article}
\usepackage{amssymb}
\usepackage{curves}
\catcode`\@=11
\def\marginnote#1{}

\newcount\hour
\newcount\minute
\newtoks\amorpm
\hour=\time\divide\hour by60 \minute=\time{\multiply\hour by60
\global\advance\minute by-\hour}
\edef\standardtime{{\ifnum\hour<12 \global\amorpm={am}%
        \else\global\amorpm={pm}\advance\hour by-12 \fi
        \ifnum\hour=0 \hour=12 \fi
        \number\hour:\ifnum\minute<10 0\fi\number\minute\the\amorpm}}
\edef\militarytime{\number\hour:\ifnum\minute<10 0\fi\number\minute}

\def\draftlabel#1{{\@bsphack\if@filesw {\let\thepage\relax
      \xdef\@gtempa{\write\@auxout{\string
          \newlabel{#1}{{\@currentlabel}{\thepage}}}}}\@gtempa \if@nobreak
    \ifvmode\nobreak\fi\fi\fi\@esphack} \gdef\@eqnlabel{#1}}
    \def\@eqnlabel{}
\def\@vacuum{}
\def\draftmarginnote#1{\marginpar{\raggedright\scriptsize\tt#1}}

\def\draft{
  \oddsidemargin -.5truein
  \def\@oddfoot{\footnotesize \sl preliminary draft \hfil
    \rm\thepage\hfil\sl\today\quad\militarytime}
  \let\@evenfoot\@oddfoot \overfullrule 3pt
    \let\label=\draftlabel
    \let\marginnote=\draftmarginnote
  \def\@eqnnum{(\theequation)\rlap{\kern\marginparsep\tt\@eqnlabel}%
    \global\let\@eqnlabel\@vacuum}

  }
\voffset= - 1.2in
\hoffset= - 1.0in         % switch off for draft style
\def\be{\begin{equation}}
\def\ee{\end{equation}}
\def\bea{\begin{eqnarray}}
\def\eea{\end{eqnarray}}
\def\<{\langle}
\def\>{\rangle}

\def\res{{{\rm res}}}

\def\d{\partial}
\def\N2{${\cal N}=2$}

\def\tr{{\mathrm{tr\,}}}

\def\1N{${\cal N}=1$}
\def\4N{${\cal N}=4$}

\def\e{{\,\rm e}\,}

\def\bea{\begin{eqnarray}}
\def\eea{\end{eqnarray}}
\def\bqa{\begin{eqnarray}}
\def\eqa{\end{eqnarray}}

\def\beq{\begin{equation}}
\def\eeq{\end{equation}}
\def\ba{\beq\begin{array}{c}}
\def\ea{\end{array}\eeq}
\def\be{\beq}
\def\ee{\eeq}

\let\text=\mathrm

\let\wht=\widehat
\let\ov=\overline

\newcommand\theTag[1]{(\ref{#1})}
\def\e{e}

\def\beq{\begin{equation}}
\def\eeq{\end{equation}}
\def\bea{\begin{eqnarray}}
\def\eea{\end{eqnarray}}

\newcommand{\rf}[1]{(\ref{#1})}

\renewcommand{\d}{{{\partial}}}

\renewcommand{\<}{\langle}
\renewcommand{\>}{\rangle}

\def\2{{1\over 2}}

\def\d{\partial}

\def\â{$\tau$}

\newcommand{\cpict}[3]{
\dimen1=#1\advance\dimen1 by-\hsize\divide\dimen1 by-2 \vtop to #2{
\noindent\hskip\dimen1{\special{em:graph #3.bmp}} \vfil}\hskip-2cm }

\newcommand{\dV}{\frac{\partial}{\partial V(p)}}
\newcommand{\cI}{\oint_{\cal C_{\cal D}}\frac{d\lambda}{2\pi i}}
\newcommand{\Vp}{V^{\prime}}

\newcommand{\ty}{{\tilde y}}

\newcommand{\iint}{\int\!\!\int}

\let\@@savethanks\thanks
\def\thanks#1{\gdef\thefootnote{\alph{footnote}}\@@savethanks{#1}}

\title{{\bf Matrix eigenvalue model: Feynman graph technique for all
genera } \vspace{.5cm}}
\author{{\bf L. Chekhov}\thanks{E-mail: \ chekhov@mi.ras.ru}
\date{ } \\ {\small
{\it Steklov Mathematical Institute, ITEP}, and {\it Laboratoire Poncelet, Moscow, Russia}}\\
and\\
{\bf B. Eynard}\footnote{E-mail: \ eynard@saclay.cea.fr }\ \
\date{ } \\
{\small {\it SPhT, CEA, Saclay, France} }}

\begin{document}
\maketitle

\vspace{-10.5cm}

\begin{center}
\hfill SPhT-T06/017\\
\hfill ITEP/TH-06/06\\
%\hfill hep-th/0604014
\end{center}

\vspace{7.5cm}

\begin{abstract}
We present the diagrammatic technique for calculating the free
energy of the matrix eigenvalue model (the model with arbitrary
power $\beta$ by the Vandermonde determinant)
to all orders of $1/N$ expansion in the
case where the limiting eigenvalue distribution spans arbitrary (but
fixed) number of disjoint intervals (curves).
\end{abstract}
%\tableofcontents
\def\thefootnote{\arabic{footnote}}

\section{Introduction\label{s:in}}

Exact solutions of various matrix models have been considerably put forward recently. The
progress in finding asymptotic expansions is mostly due to a geometrization of the picture.
Attained first for a mere large-$N$ limit of the Hermitian one matrix model (1MM) \cite{Kri}, \cite{DV}
 \cite{ChMir}, \cite{ChMarMirVas}, it was almost simultaneously put forward for solutions
of two-matrix Hermitian matrix model (2MM) \cite{eynm2m}, \cite{eynm2mg1}, \cite{KazMar}, \cite{MarcoF}, \cite{Bertola}.
Combined with the technique
of the loop equation \cite{loop}, which is the generating function for Virasoro conditions in
matrix models \cite{Virasoro}, it permitted to generalize the moment technique of \cite{ACKM} to finding
the subleading order correction to the 1MM free energy first in the two-cut case~\cite{Ak96} and then in the multicut
case~\cite{Kos},~\cite{DST},~\cite{Chekh}. Almost simultaneously, the same correction was found in the
2MM case \cite{EKK} and in the case of normal matrices (see~\cite{WZ1},~\cite{WZ2}).

Next step in constructing asymptotic expansions pertains to introducing the {\em diagrammatic technique}
describing terms of expansion in the way very similar to the one in (quantum) field theories: the
expansion order is related to the number of loops in the diagrams, and $n$-point correlation functions
as well as the free energy itself are presented by a finite sum of diagrams in each given
order of the expansion. This technique, first elaborated in~\cite{eynloop1mat} for correlation functions
in the 1MM case, was then developed for pure and mixed correlation functions in the 2MM case~\cite{EOtrmixte},~\cite{EyOran}
and then turned into a complete solution for the free energy first in the 1MM case~\cite{ChEy},
and then, eventually, in the 2MM case~\cite{ChEyOr}.

We expect that generalizations of the diagrammatic technique are a very power tool for investigating
various matrix-model-like problems. We address the eigenvalue model, which describes
a gas of $N$ particles dwelling in the potential field and having anyonic statistics expressed by a power
of the Vandermonde determinant in coordinates of the particles. This is the Dyson (or Laughlin) gas
system, and we investigate a one-dimensional case of this system in this paper. (The first investigation
in two-dimensional case was done recently in~\cite{WZ3} where two first subleading order corrections were
derived but without involving a diagrammatic technique.) We elaborate the diagrammatic technique for
consistent calculation of corrections of all orders and find separately the only correction term that
cannot be produced from this technique.

The paper is organized as follows: in Sec.~\ref{s:1MM}, we formulate the problem and present the loop
equation together with necessary definitions. In Sec.~\ref{s:diagram}, we introduce elements of our
diagrammatic technique for calculating resolvents (loop means), and we summarize this technique in
Sec.~\ref{s:rules}. In the next section~\ref{s:freeBeta}, we invert the action of the loop insertion operator
and obtain the free energy of the model, presenting first few corrections. This enables us to obtain all
correction terms except two: the first one is just a subleading term in the 1MM case (see~\cite{Chekh}), while
the second term has not been known previously. We calculate it separately in Sec.~\ref{s:F02}.

\section{Eigenvalue models in $1/N$-expansion \label{s:1MM}}

Our aim is to show how the technique of Feynman graph expansion elaborated in the
case of Hermitian one-matrix model~\cite{eynloop1mat},~\cite{ChEy} can be applied to solving the
(formal) eigenvalue model with the action
\be
%\int_{t_0 N\times t_0 N}DX\, \e^{-N\tr V(X)}=\e^{\cal F},
\int\prod_{i=1}^N dx_i\, \Delta^{2\beta}(x)\e^{-{N\beta\over t_0}\,\sum_{i=1}^N V(x_i)}=\e^{-\cal F},
\label{X.1}
\ee
where $V(x)=\sum_{n \geq 0}^{}t_nx^n$ and
\be
\hbar = {t_0\over  N\sqrt\beta}
\ee
is a formal expansion parameter. The integration in (\ref{X.1}) goes
over $N$ variables $x_i$ having the sense of eigenvalues of the
Hermitian matrices for $\beta=1$, orthogonal matrices for $\beta=1/2$,
and symplectic matrices for $\beta=2$. In what follows, we set $\beta$ to be
arbitrary positive number. The
integration may go over curves in the complex plane of each
of $N$ variables $x_i$. For $\beta\ne1$, no topological expansion
in even powers of $\hbar$ exists and we rather have the expansion in all
integer powers of $\hbar$.
Customarily, $t_0=\hbar N$ is the scaled
number of eigenvalues. We also assume
the potential $V(p)$ to be a polynomial
of the fixed degree $m+1$, or, in a more general setting, it suffices
to demand the derivative $V'(p)$ to be a rational function~\cite{eylooprat}.

The averages corresponding to partition function~\theTag{X.1} are
defined in a standard way:
\beq
\bigl\langle Q(X)\bigr\rangle=
\frac1Z\int_{N\times N}DX\,Q(X)\,\exp\left(-N{\sqrt\beta\over \hbar}\tr V(X)\right),
\label{X.3*}
\eeq
and we introduce their formal
generating functionals: the one-point resolvent
\beq
W(p)=%\frac{1}{M}
\hbar
\sqrt\beta\,\sum_{k=0}^{\infty}
\frac{\langle\tr X^{k}\rangle}{p^{k+1}}
\label{X.3}
\eeq
as well as the $s$-point resolvents $(s\geq2)$
\bea
W(p_1,\dots,p_s)
&=&
%M^{s-2}
\beta\,\left(\hbar\over\sqrt\beta\right)^{2-s}\,
\sum_{k_1,\dots,k_s=1}^{\infty}
\frac{\langle\tr X^{k_1}\cdots\tr X^{k_s}\rangle_{\mathrm{conn}}}
{p_1^{k_1+1}\cdots p_s^{k_s+1}} \cr
%M^{s-2}
&=&
\beta\,\left(\hbar\over\sqrt\beta\right)^{2-s}\,
\left\langle\tr\frac{1}{p_1-X}\cdots
\tr\frac{1}{p_s-X}\right\rangle_{\mathrm{conn}}
\label{X.4}
\eea
where the subscript ``$\mathrm{conn}$" pertains to the connected
part.

These resolvents are obtained from the free energy ${\cal F}$ through the
action
\bea
W(p_1,\dots,p_s)&=&-\hbar^2\frac{\d}{\d V(p_s)}\frac{\partial}{\partial V(p_{s-1})}\cdots
\frac{\partial {\cal F}}{\partial V(p_1)}=\nonumber
\\
&=&\frac{\partial }{\partial V(p_s)}\frac{\partial }{\partial V(p_{s-1})}\cdots
\frac{\partial }{\partial V(p_2)}W(p_1),
\label{X.5}
\eea
of the loop insertion operator
\beq
\frac{\partial }{\partial V(p)}\equiv
-\sum_{j=1}^{\infty}\frac{1}{p^{j+1}}\frac{\d}{\d t_{j}}.
\label{X.6}
\eeq
Therefore, if one knows exactly the one-point resolvent for arbitrary
potential, all multi-point resolvents can be calculated by induction.
In the above normalization, the $\hbar$-expansion has the form
\beq
W(p_1,\dots,p_s)=\sum_{r=0}^{\infty}
%\frac{1}{M^{2g}}
\hbar^{r}
W_{r/2}(p_1,\dots,p_s),\quad s\geq1,
\label{X.7}
\eeq
where it is customarily assumed that it corresponds in a vague sense
to the genus expansion in the usual Hermitian models with possible half-integer
contributions. It is often written as a sum over all---integer
and half-integer---$g\equiv r/2$.

The first in the chain of the loop equations of the eigenvalue model (\ref{X.1}) is
\beq
\oint_{{\cal C}_{\cal D}}\frac{d\omega}{2\pi i}\frac{V'(\omega)}{p-\omega}W(\omega)=
W(p)^2+\hbar\left(\sqrt{\beta}-\sqrt{\beta^{-1}}\right)W'(p)
%\frac{1}{M^2}
+\hbar^2 W(p,p).
\label{X.8}
\eeq
Here and hereafter, ${\cal C}_{\cal D}$~is a contour encircling clockwise
all singular points (cuts)
of $W(\omega)$, but not the point
$\omega=p$; this contour integration acts as the projection operator
extracting negative part of $V'(p)W(p)$. Using Eq.~\theTag{X.5}, one can express
the third second term in the r.h.s.\ of loop equation~\theTag{X.8} through
$W(p)$, and Eq.~\theTag{X.8} becomes an equation on
one-point resolvent \rf{X.3}.

The $\beta$-dependence enters (\ref{X.8}) only through the combination
\be
\gamma=\sqrt{\beta}-\sqrt{\beta^{-1}},
\label{X.8*}
\ee
and, assuming $\beta\sim O(1)$, we have the free energy expansion
of the form
\be
{\cal F}\equiv {\cal F}(\hbar, \gamma, t_0, t_1, t_2, \dots)
=\sum_{k=0}^{\infty}\sum_{l=0}^{\infty}{\hbar}^{2k+l-2}\gamma^l{\cal F}_{k,l}.
\label{X.2}
\ee

Substituting expansion~\theTag{X.7} in Eq.~\theTag{X.8}, we find
that $W_g(p)$ for $g\geq1/2$ satisfy the equation
\beq
\left(\widehat{K}-2W_{0}(p)\right)W_g(p)=\sum_{g'=1/2}^{g-1/2}
W_{g'}(p)W_{g-g'}(p)+\frac{\partial }{\partial V(p)}W_{g-1}(p)+\gamma \frac{\d}{\d p}W_{g-1/2}(p),
\label{X.9}
\eeq
where $\widehat{K}$~is the linear integral operator
\beq
\widehat{K}f(p)\equiv-\oint_{{\cal C}_{\cal D}}\frac{d\xi}{2\pi i}
\frac{V'(\xi)}{p-\xi}f(\xi).
\label{X.10}
\eeq
In Eq.~\theTag{X.9}, $W_g(p)$ is expressed through only the
$W_{g_i}(p)$ for which $g_i<g$. This fact permits
developing the iterative procedure.

In analogy with (\ref{X.2}), it is  convenient to expand multiresolvents $W_g(\cdot)$ in $\gamma$:
\be
W_g(p_1,\dots,p_s) = \sum_{k=0}^{[g]}\,\gamma^{2g-2k}\,W_{k,2g-2k}(p_1,\dots,p_s)
\label{X.11}
\ee
Then, obviously, (\ref{X.9}) becomes
\be
\left(\widehat{K}-2W_{0,0}(p)\right)W_{k,l}(p)=\sum_{k_1\ge0,l_1\ge0\atop k_1+l_1>0}
W_{k_1,l_1}(p)W_{k-k_1,l-l_1}(p)+\frac{\partial }{\partial V(p)}W_{k-1,l}(p)+\gamma \frac{\d}{\d p}W_{k,l-1}(p).
\label{X.12}
\ee

The form of loop equation (\ref{X.8}) is based exclusively on the
reparameterization invariance of the matrix integral, which retains
independently on the details of eigenvalue density distribution.
We assume that as $N\to\infty$, the eigenvalues fill in some segments
in complex plane, dependently on the shape of potential $V(X)$. For
polynomial potentials, the number of segments is finite and
the contour ${\cal C_{\cal D}}$ of integration in (\ref{X.9})
encircles a finite number~$n$ of disjoint intervals
\be
{\cal D} \equiv \bigcup_{i=1}^n [\mu_{2i-1},\mu_{2i}],
\quad \mu_1< \mu_2< \ldots < \mu_{2n}.
\label{defcalD}
\ee
Recall that
\be
W_{k,l}(p)|_{p\to\infty} = \frac{t_0}{p}\delta_{k,0}\delta_{l,0}+O({1}/{p^2}),
\label{Winf}
\ee
all $W_{k,l}(p)$ are total derivatives,
\be
\label{total}
W_{k,l}(p)=- \dV {\cal F}_{k,l},\quad k,l\ge 0.
\ee

The solution in the large-$N$ limit coincides with the solution of the
Hermitian one-matrix model and satisfies the equation
(in fact, the above normalization was chosen in a way to ensure the
coincidence of this equation with the one in the 1MM case)
\be
\cI \frac{V^{\prime}(\lambda)}{p-\lambda} W_{0}(\lambda) = (W_{0}(p))^2.
\label{plan}
\ee
Actually, $W_0(p)$ must be $W_{0,0}(p)$ in our notation; we however preserve the old notation assuming
this identification in what follows (for shortening the presentation).

Recall the solution of Eq. (\ref{plan}).
Deforming the contour in Eq.~(\ref{plan}) to infinity, we obtain
\be
(W_{0}(p))^2  = \Vp(p) W_0(p)
 +\oint_{\cal C_{\infty}}\frac{d\lambda}{2\pi i} \frac{\Vp(\lambda)}{p-\lambda}W_0(\lambda),
 \label{*loop1*}
\ee
where the last term in the r.h.s. is
a polynomial $P_{m-1}$ of degree $m-1$, and the solution to \rf{*loop1*}
is then
\be
W_0(p) = \frac{1}{2}\Vp(p) - \frac{1}{2}\sqrt{\Vp(p)^2+4P_{m-1}(p)}
\equiv \frac{1}{2}\Vp(p) - y(p),
 \label{*loop2*}
\ee
where the minus sign is chosen in order to fulfill asymptotic condition (\ref{Winf})
whereas the function $y(p)$ is defined as follows.
For a polynomial potential $V$ of degree $m+1$, the resolvent $W_0(p)$ is
a function on complex plane with $n\leq m$ cuts, or on a hyperelliptic curve
$y^2 = \Vp(p)^2+4P_{m-1}(p)$ of genus $g=n-1$.
For generic potential $V(X)$ with $m\to\infty$, this curve
may have an infinite degree, but we can still consider solutions with a finite genus, where a fixed number $n$ of cuts are filled by eigenvalues.
For this, we separate the smooth part of the curve introducing
\be
\label{ty}
y(p)\equiv M(p)\ty(p), \quad \hbox{and} \quad
{\ty}^2(p)\equiv\prod\nolimits_{\alpha=1}^{2n}(p-\mu_\alpha)
\ee
with all branching points $\mu_\alpha$ distinct. The variable $\ty$ defines therefore
the new, reduced Riemann surface, which plays a fundamental role in our construction.
In what follows, we still assume
$M(p)$ to be a polynomial of degree $m-n$, keeping in mind that $n$ is always
finite and fixed, while $m\geq n$ can be chosen arbitrarily large.

From now on, we distinguish between images of the infinity at two sheets---physical
and unphysical---of hyperelliptic Riemann surface (\ref{ty}) respectively denoting them
$\infty_+$ and $\infty_-$. We often distinguish between variables of the physical and
unphysical sheets placing the bar over the latter.
By convention, we set $\ty|_{p\to\infty_+}\sim p^{n}$, and
$M(p)$ is then\footnote{By a standard convention, $\res_{\infty}dx/x=-1$, and the
direction of the integration contour at the infinity point therefore coincides with the
direction of contour for integrals over $\cal C_{\cal D}$ and over the set of $A$-cycles, see below.}
\be
M(p) =-\frac12 \res_{\infty_+} {dw} \frac{\Vp(w)}{(w-p)\ty(w)}.
\label{M}
\ee
Inserting this solution in Eq. (\ref{*loop2*}) and deforming
the contour back, we obtain the planar one-point resolvent
with an $n$-cut structure,
\be
W_0(p) = \frac{1}{2}\cI \frac{\Vp(\lambda)}{p-\lambda}\frac{\ty(p)}{\ty(\lambda)},
\quad p\not\in{\cal D}.
\label{W0}
\ee

Let us now discuss the parameter counting.
We introduce the {\em filling fractions}
\be
\label{Sfr}
S_i =
\oint_{A_i}\frac{d\lambda}{2\pi i}\,y(\lambda)=\oint_{A_i}\frac{d\lambda}{2\pi i}M(\lambda)\ty(\lambda),
\ee
where $A_i$, $i=1,\dots,n-1$ is the basis of $A$-cycles on the hyperelliptic
Riemann surface \rf{ty} (we may conveniently choose them to be the first $n-1$ cuts).
Adding the (normalized) total number of eigenvalues
\be
\label{t0}
t_0 = \oint_{\cal C_{\cal D}}y(\lambda) \frac {d\lambda}{2\pi i} = \res_{\infty_+}\, y(\lambda)d\lambda
\ee
to the set of $S_i$, we obtain $n$ parameters,
to which we add, following (\ref{Winf}), the asymptotic conditions
\be
-t_0\delta_{k,n}  =  \frac{1}{2}
 \cI \frac{\lambda^k V^{\prime}(\lambda)}{\ty(\lambda)}, \quad k=0,\ldots,n.
 \label{Rand1}
\ee

In this article  we consider filling fractions \rf{Sfr}
as {\it independent\/} parameters of the theory. We need
this assumption to interpret random matrix integrals as generating functions of discrete surfaces.
Other assumptions are possible, but we do not consider them in this paper.
In other words, we consider only the perturbative part of the matrix integral, and the filling fractions are fixed because  the
jumps between different cuts are non-perturbative corrections in $\hbar$.
In particular, this imposes restrictions
\be
\label{dVSi}
\dV S_i=0,\quad i=1,\dots,n-1,\qquad \dV t_0=0.
\ee
That implies that, for $k+l>0$:
\beq\label{vanishAcycle}
\oint_{A_i} W_{k,l}(\xi,p_2,\dots,p_s) = 0
\eeq

In addition, we impose another assumption.
The zeroes $b_j$ of $M(p)$ are called double points. In some sense, they can be considered as cuts of vanishing size.
We require that those degenerate cuts contain no eigenvalue to any order in the $\hbar$ expansion. We therefore demand
\beq
\oint_{{\cal C}_{b_j}} W_{k,l}(\xi,p_2,\dots,p_s) = 0
\eeq
for any contour which encircles a zero of $M$.

This assumption, together with loop equation \rf{X.12}, suffices for proving that for every $k,l,s$,
except $(k,l,s)=(0,0,1)$ and $(k,l,s)=(0,0,2)$, the function $W_{k,l}(p_1,\dots,p_s)$ has singularities
in the physical sheet only at the branch points $\mu_\alpha$. In particular,
it has no singularities at the double points in the physical sheet.

\section{Calculating resolvents. Diagrammatic technique  \label{s:diagram}}

In this section, we derive the diagrammatic technique for model (\ref{X.1}),
which is a generalization of technique in~\cite{eynloop1mat},~\cite{ChEy}.
Our main goal
is to invert loop equation (\ref{X.9}) to obtain the expression for $W_k(p)$
for any $k\ge1/2$.

\subsection{A piece of Riemann geometry}\label{ss:B(P,Q)}

The main notion is again the
{\it Bergmann kernel},\footnote{It is a double derivative of the logarithm of the Prime form.}
which is the unique bi-differential on a Riemann
surface~$\Sigma_g$ that is symmetrical in its arguments $P,Q\in \Sigma_g$
and has the only singularity (a double pole) at the coinciding arguments where, in any local coordinate $\tau$, it has the
behavior (see~\cite{Fay},~\cite{Farkas})
\be
B(P,Q)=\left(\frac{1}{(\tau(P)-\tau(Q))^2}+\frac16S_B(P)+o(1)\right)d\tau(P)d\tau(Q),
\label{*Bergmann*}
\ee
with $S_B(P)$ the Bergmann projective connection associated to the local coordinate $\tau$.
We fix the normalization claiming
vanishing all the integrals over $A$-cycles of $B(P,Q)$:
\be
\oint_{A_i}B(P,Q)=0,\ \hbox{for}\ i=1,\dots,g.
\label{*vanish*}
\ee
We then have the standard Rauch variational formulas relating $B(P,Q)$ with other objects on
a (general, not necessarily hyperelliptic) Riemann surface:
\be
\frac{\d}{\d\mu_\alpha}B(P,Q)= 2 B(P,[\mu_\alpha])B([\mu_\alpha],Q),
\label{dB}
\ee
and
\be
\oint_{B_i}B(P,Q)=2\pi i\, dw_i(P),
\label{B-int}
\ee
where $\mu_\alpha$ is any simple branching point of the complex structure. Then, by definition,
in the vicinity of $\mu_\alpha$,
\be
B(P,Q)|_{Q\to\mu_\alpha}=B(P,[\mu_\alpha])\left(\frac{dq}{\sqrt{q-\mu_\alpha}}+O(\sqrt{q-\mu_\alpha})dq\right),
\label{Y.1}
\ee
and $dw_i(P)$ are canonically normalized holomorphic differentials:
\be
\oint_{A_j}dw_i(P)=\delta_{ij}.
\label{dw}
\ee

Besides these formulas, we need another, rather obvious, relation: for any meromorphic function $f$ on the curve, we have:
\be
df(P) = \frac{1}{2\pi i}\oint_{{\cal C}_P}B(P,\xi)\,f(\xi)
%\frac{\d}{\d P}{B(P,Q)\over dP}=\frac{1}{2\pi i}\oint_{{\cal C}_P}{B(P,\xi)B(\xi,Q)\over dP\,d\xi},
\label{der}
\ee
where the contour ${\cal C}_P$ encircles the point $P$ only, and not the poles of $f$.

We also introduce the 1-form $dE_{Q,q_0}(P)$, which is the primitive of $B(P,Q)$:
\be
dE_{Q,q_0}(P)=\int_{q_0}^QB(P,\xi),\qquad dE_{Q,q_0}(P)|_{P\to Q}=\frac{d\tau(P)}{\tau(P)-\tau(Q)}+\hbox{finite\,}.
\label{dE}
\ee
Then, obviously,
\be
\oint_{A_i}dE_{Q,q_0}(P)=0.
\label{dEA}
\ee
The form $dE_{Q,q_0}(P)$ is single-valued w.r.t. $P$ on the Riemann surface and multiple-valued w.r.t. the
variable $Q$: from (\ref{B-int}),
\bea
&&dE_{Q+\oint_{B_i},q_0}(P)=2\pi i dw_i(P)+dE_{Q,q_0}(P),
\nonumber
\\
&&dE_{Q+\oint_{A_i},q_0}(P)=dE_{Q,q_0}(P).
\nonumber
\eea
where the reference point $q_0$ will happen to play no role and it must disappear eventually from all formulae.
Technically it is convenient to have $q_0$ in the non-physical sheet.

We can now express the 2-point resolvent
$W_0(p,q)$ in terms of $B(P,Q)$. We let $p$ and $\ov p$ denote the complex coordinates of
points on the respective physical and unphysical sheets. Then,
$$
dp\, dq \,\frac{\d V'(p)}{\d V(q)}=-B(p,q)-B(p,\ov q)=-\frac{dp\,dq}{(p-q)^2}
$$
since it has double poles with unit quadratic residues at $p=q$ and $p=\ov q$.
The 2-point resolvent (\ref{*loop2*}) is nonsingular at coinciding points; therefore,
\be
dp\,dq\,\frac{\d y(p)}{\d V(q)}=-\frac12 (B(p,q)-B(p,\ov q)),
\label{dVy}
\ee
and
\be
W_0(p,q)=-{B(p,\ov q)\over dp\,dq}.
\label{W2}
\ee

\subsection{Inverting the operator $\wht K-2W_0(p)$}\label{ss:dE}

We can determine corrections in $\hbar$
iteratively by inverting
loop equation (\ref{X.9}). All
multi-point resolvents of the same order can be obtained from $W_g(p)$ merely
by applying the loop insertion operator $\dV$.

In the 1MM case, a natural restriction imposed on the free energy is that
all the higher free energy terms $F_g$ must depend only on $\mu_\alpha$
and a {\em finite} number of the moments $M_\alpha^{(k)}$, which are derivatives
of $(k-1)$th order of the polynomial $M(p)$ at branching points, allowed
no freedom of adding the terms depending only on $t_0$ and $S_i$ to ${\cal F}_g$.
In model (\ref{X.1}), such a restriction cannot be literally imposed, as we demonstrate
below, and we instead claim that
\be
\frac{\d {\cal F}}{\d S_i}=\oint_{B_i}\frac{\d {\cal F}}{\d V(\xi)}d\xi,
\label{Z.5}
\ee
which was a consequence of locality in 1MM and can be taken as a defining relation in
the $\beta$-model.

\medskip

The first step is to find the inverse of the operator $\wht K-2W_0(p)$.
It was found in \cite{eynloop1mat}, that if $f(p)$ is a function whose only singularities in the physical sheet
are cuts along ${\cal D}$, which vanishes at $\infty$ like $O(1/p^2)$ in the physical sheet
and has vanishing A-cycle integrals, then, having $dE_{q,\bar q}(p)=dE_{q, q_0}(p)-dE_{\bar q,q_0}(p)$,
\be
\widehat{d{\cal E}}_{p,q}(\wht{K}-2W_0(q))f(q)\equiv
\frac{1}{2\pi i}\, \oint_{{\cal C}_{\cal D}} \frac{dE_{q,\bar q}(p)\,dq}{2y(q)} (\wht K-2W_0(q)).f(q) = f(p)\,dp,
\label{Y.2}
\ee
where $y(q) = {V'(q)\over 2} - W_0(q)$ is the function introduced in \rf{*loop2*} and integration
contour lies in the physical sheet. Indeed, in the standard notation for the projections $[Q(q)]_\pm$ that
segregate the respective polynomial and regular parts of $Q(q)$ as $q\to\infty$, we have
$(\wht K-2W_0(q)).f(q) =2[y(q)f(q)]_-= 2y(q) f(q) - P(q)$ where $P(q)$ is a polynomial $[y(q)f(q)]_+$
of degree $\deg V'-2$, and thus:
\be
 \frac{1}{2\pi i}\, \oint_{{\cal C}_{\cal D}} \frac{dE_{q,\bar q}(p)\,dq}{2y(q)} (\wht K-2W_0(q)).f(q)
= \frac{1}{2\pi i}\, \oint_{{\cal C}_{\cal D}} {dE_{q,\bar q}(p)\,dq}\, \left(f(q)-{P(q)\over 2y(q)}\right) .
\ee
Let us compute the part of the integral involving the polynomial $P(q)$:
\bea
 \oint_{{\cal C}_{\cal D}} {dE_{q,\bar q}(p)\,dq}\, {P(q)\over 2y(q)}
&=&  \oint_{\bar{{\cal C}_{\cal D}}} {dE_{q,\bar q}(p)\,dq}\, {P(q)\over y(q)}   \cr
&=& - \oint_{{\cal C}_{\cal D}} {dE_{q,\bar q}(p)\,dq}\, {P(q)\over y(q)} +
2i\pi \sum_\alpha \mathop{{\rm Res}}_{\mu_\alpha} {dE_{q,\bar q}(p)\,dq}\, {P(q)\over y(q)} \cr
&=& - \oint_{{\cal C}_{\cal D}} {dE_{q,\bar q}(p)\,dq}\, {P(q)\over y(q)}  \cr
&=& 0.
\eea
The first equality is obtained by changing the name of the variable $q\to \bar q$ with
accounting for $P(\bar q)=P(q)$, $y(\bar q)=-y(q)$. The second equality comes from deforming the contour
$\bar {\cal C}_{\cal D}$ to $-{\cal C}_{\cal D}$ picking residues at branch points.
The third equality holds because $P(q)$ is a polynomial and thus has no singularity at branch points, whereas
zeros of $y(q)$ are canceled by those of $dE_{q,\bar q}(p)$, so the residues vanish.

Therefore we have:
\bea
 \frac{1}{2\pi i}\, \oint_{{\cal C}_{\cal D}} \frac{dE_{q,\bar q}(p)\,dq}{2y(q)} (\wht K-2W_0(q)).f(q)
&=& \frac{1}{2\pi i}\, \oint_{{\cal C}_{\cal D}} {dE_{q,q_0}(p)\,dq}\, f(q) - \frac{1}{2\pi i}\, \oint_{{\cal C}_{\cal D}} {dE_{\bar q,q_0}(p)\,dq}\, f(q) \cr
%&=& \frac{1}{2\pi i}\, \oint_{{\cal C}_{\cal D}} {dE_{q,q_0}(p)\,dq}\, f(q) - \frac{1}{2\pi i}\, \oint_{\bar{{\cal C}_{\cal D}}} {dE_{q,q_0}(p)\,dq}\, f(\bar q) \cr
&=& \frac{1}{2\pi i}\, \oint_{{\cal C}_{\cal D}} {dE_{q,q_0}(p)\,dq}\, f(q)  \cr
&=& \frac{1}{2\pi i}\, \oint_{{\cal C}'_{\cal D}} {(dE_{q,q_0}(p)+dw(p))\,dq}\, f(q) \cr
&=& \frac{1}{2\pi i}\, \oint_{{\cal C}'_{\cal D}} dE_{q,q_0}(p)\,dq\, f(q) \cr
&=& - \mathop{{\rm Res}}_{q\to p} dE_{q,q_0}(p)\,dq\, f(q) \cr
&=& f(p)\,dp .
\eea
The second equality holds because $dE_{\bar q,q_0}(p)$ has no singularity at $q\to p$ and we can push the integration contour for $q$ to
infinity (in the physical sheet), which gives zero.
In the third equality, the contour ${\cal C}'_{\cal D}$ is a contour which encloses the $A$-cycles and ${\cal D}$. When we cross the cycle $A_i$,
$dE_{q,q_0}(p)$ jumps by the corresponding holomorphic differential $dw_i(p)$.
The fourth equality holds because $dw_i(p)$ is independent of $q$, whereas the integral of $f(q)$ along
any $A-$cycle vanishes by our assumption \rf{vanishAcycle} (this is nothing but the Riemann bilinear identity).
Then the contour ${\cal C}'_{\cal D}$ is deformed in the physical sheet, into a contour which encloses only the point $p$
(recall that we have assumed that $f(q)$
has no singularity in the physical sheet, and vanishes like $O(1/q^2)$ at $\infty$),
and the final result comes from the fact that $dE$ has a simple pole (see eq. \rf{dE}).

This proves that for all functions of the type $W_{k,l}$,
the integration $\frac{1}{2\pi i}\, \oint_{{\cal C}_{\cal D}} \frac{dE_{q,\bar q}(p)\,dq}{2y(q)}$ acts like the inverse of $\wht K-2W_0(q)$.

\smallskip

Notice that if $f(q)$ is  a function which has poles only at branch points $\mu_\alpha$
or at double points $b_j$, by moving the integration contours we have:
\beq
 \frac{1}{2\pi i}\, \oint_{{\cal C}_{\cal D}} \frac{dE_{q,\bar q}(p)\,dq}{2y(q)}\, f(q)
= \sum_\alpha  \mathop{{\rm Res}}_{q\to \mu_\alpha} \frac{dE_{q,q_0}(p)\,dq}{2y(q)}\, f(q)  +
\sum_j  \mathop{{\rm Res}}_{q\to b_j} \frac{dE_{q,q_0}(p)\,dq}{2y(q)}\, (f(q)-f(\bar q)).
\eeq

\smallskip

This relation provided a basis for the diagrammatic representation for resolvents in 1MM~\cite{eynloop1mat},
and we show that it works in the case of $\beta$-model as well.
Let us represent the form $dE_{q,\bar q}(p)$ as the vector directed
from $p$ to $q$, the three-point vertex as the dot at which
we assume the integration over $q$, \ $\bullet\equiv\oint_{{\cal C}_{\cal D}}\frac{dq}{2\pi i}\frac{1}{2y(q)}$, and the
Bergmann 2-form $B(p,q)$ as a nonarrowed edge connecting points $p$ and $q$.

Let us also introduce a new propagator $dpdy(q)$ denoted by the dashed line.

The graphic representation
for a solution of (\ref{X.9}) then looks as follows.
We represent the multiresolvent $W_{g'}(p_1,\dots,p_s)$ as the block with $s$ external legs and with the index $g'$.
We also present the derivative $\frac{\d}{\d p_1}W_{g'}(p_1,\dots,p_s)$ as the block with $s+1$ external legs, one of which
is the dashed leg that starts at the same vertex as $p_1$. That is,
we obtain~(cf.~\cite{eynloop1mat})
\be
\begin{picture}(190,55)(50,10)
\thicklines
\put(35,40){\oval(20,20)}
\thinlines
\put(15,40){\line(1,0){10}}
\put(15,40){\circle*{2}}
\put(15,44){\makebox(0,0)[cb]{$p $}}
\put(35,40){\makebox(0,0)[cc]{$g$}}
\put(50,40){\makebox(0,0)[lc]{$=\sum\limits_{g'=1/2}^{g-1/2}$}}
\put(80,40){\vector(1,0){13}}
\put(80,40){\circle*{2}}
\put(95,40){\circle*{4}}
\put(80,44){\makebox(0,0)[cb]{$p$}}
\put(93,44){\makebox(0,0)[cb]{$q$}}
\put(95,40){\line(1,1){7}}
\put(95,40){\line(1,-1){7}}
\thicklines
\put(109,54){\oval(20,20)}
\put(109,26){\oval(20,20)}
\put(109,54){\makebox(0,0)[cc]{$g{-}g'$}}
\put(109,26){\makebox(0,0)[cc]{$g'$}}
\thinlines
\put(120,40){\makebox(0,0)[lc]{$+$}}
\put(130,40){\vector(1,0){13}}
\put(130,40){\circle*{2}}
\put(145,40){\circle*{4}}
\put(130,44){\makebox(0,0)[cb]{$p$}}
\put(143,44){\makebox(0,0)[cb]{$q$}}
\qbezier(145,40)(150,49)(157,49)
\qbezier(145,40)(150,31)(157,31)
\thicklines
\put(166,40){\oval(25,25)}
\put(166,40){\makebox(0,0)[cc]{$g{-}1$}}
\thinlines
\put(190,40){\makebox(0,0)[lc]{$+\ \gamma$}}
\put(205,40){\vector(1,0){13}}
\put(205,40){\circle*{2}}
\put(220,40){\circle*{4}}
\put(205,44){\makebox(0,0)[cb]{$p$}}
\put(218,44){\makebox(0,0)[cb]{$q$}}
\curve(220,40, 224,35, 232,31)
\curvedashes[0.5mm]{2,2}
\curve(220,40, 224,45, 232,49)
%\put(220,40){\line(1,1){9}}
%\put(220,40){\line(1,-1){9}}
\thicklines
\put(241,40){\oval(25,25)}
\put(241,40){\makebox(0,0)[cc]{$g{-}\frac12$}}
\put(260,40){\makebox(0,0)[cc]{$,$}}
\end{picture}
\label{Bertrand}
\ee
which provides the diagrammatic representation for $W_k(p_1,\dots,p_s)$.

Recall the diagrammatic formulation of 1MM ($\gamma=0$). There
the multiresolvent
$W_{k,0}(p_1,\dots,p_s)$ can be presented as a finite sum of all possible connected
graphs with $k$ loops and $s$ external legs and with only three-valent internal vertices
(the total number of edges is then $2s+3k-3$, and we assume $s\ge1$ for $k\ge1$ and $s\ge3$ for $k=0$)
and such that in each graph we segregate a maximal rooted tree subgraph
with all arrows directed from the root. This subtree comprises exactly $2k+s-2$ arrowed edges.
We then choose one of the external legs, say, $p_1$
(the choice is arbitrary due to the symmetry of $W_{k,0}(p_1,\dots,p_s)$), to be
the root vertex the tree starts with; for each three-valent
vertex there must exist exactly one incoming edge of the tree subgraph. All external edges (except the
root edge) are lines corresponding to $B(p,q)$ and are therefore  nonarrowed.
This subtree therefore establishes a partial ordering of vertices: we say that
vertex~A {\it precedes\/} vertex~B if there exists
a directed path in the subtree from A to~B. Internal nonarrowed edges are again $B(r,q)$ but we allow
only those nonarrowed edges for which the endpoints, $r$ and $q$, are {\em comparable}. If $r=q$, then, for the
tadpole subgraph, we set $B(r,\bar r)$, where $\bar r$ is the point on the other, nonphysical sheet.
At each internal vertex, denoted by $\bullet$, we have the integration
$\oint_{{\cal C}^{(q)}_{\cal D}}\frac{dq}{2\pi i}\frac{1}{2y(q)}$, while the arrangement of the integration contours
at different internal vertices is prescribed by the arrowed subtree: the closer is a vertex to the root, the more outer
is the integration contour.\footnote{Since no propagator connects noncomparable vertices, the mutual ordering of the
corresponding integration contours is irrelevant.}

\subsection{Acting by spatial derivative}\label{ss:dx}

As a warm-up example, let us consider the action of the spatial derivative $\d/\d p_1$ on $W_{k,0}(p_1,\dots,p_s)$.
We place the starting point of the ( fictitious) dashed directed edge to the same point $p_1$ and associate
just $dx$ with this starting point. Recall that the first object (on which the derivative actually acts) is
$dE_{\eta,\bar\eta}(p_1)$, then comes the vertex with the integration $\frac{1}{2\pi i}\oint_{{\cal C}_{\cal D}^{(\eta)}}\frac{d\eta}{y(\eta)}$,
then the rest of the diagram, which we denote as $F(\eta)$. We can present the action of the derivative
via the contour integral around $p_1$ with the kernel $B(p_1,\xi)$:
\be
dp_1\,{\d\over \d {p_1}}\left(\oint_{{\cal C}_{\cal D}^{(\eta)}}\frac{dE_{\eta,\bar\eta}(p_1) d\eta}{2\pi i \ y(\eta)\, dp_1}F(\eta)\right)
= \mathop{{\rm Res}}_{\xi\to p_1} \oint_{{\cal C}_{\cal D}^{(\eta)}} \,\frac{B(p_1,\xi)dE_{\eta,\bar\eta}(\xi)}{d\xi} \,\frac{d\eta}{2\pi i \ y(\eta)}\,F(\eta),
\label{Y.3}
\ee
where $p_1$ lies outside the integration contour for $\eta$. The integral over $\xi$ is nonsingular at infinity, so we can
deform the integration contour from ${\cal C}_{p_1}$ to ${\cal C}_{\cal D}^{(\xi)}>{\cal C}_{\cal D}^{(\eta)}$\footnote{Here, as in
\cite{ChEy}, comparison of contours is equivalent to their inside/outside ordering.}.
\beq
 \oint_{{\cal C}_{p_1}}\frac{B(p_1,\xi)dE_{\eta,\bar \eta}(\xi)}{2\pi i \ d\xi} \oint_{{\cal C}_{\cal D}^{(\eta)}}\frac{d\eta}{2\pi i \ y(\eta)}F(\eta)
= - \oint_{{\cal C}_{\cal D}^{(\xi)} >{\cal C}_{\cal D}^{(\eta)}}\frac{B(p_1,\xi)dE_{\eta,\bar \eta}(\xi)}{2\pi i \ d\xi} \frac{d\eta}{2\pi i \ y(\eta)}F(\eta)
\eeq
We now push the contour for $\xi$ through the contour for $\eta$, picking residues at the poles in $\xi$, at $\xi=\eta$, $\xi=\bar\eta$
and at the branch points. We then obtain
\beq
\label{Y.4}
= - \oint_{{\cal C}_{\cal D}^{(\eta)}} \sum_\alpha \mathop{{\rm Res}}_{\xi\to \mu_\alpha}
\frac{B(p_1,\xi)dE_{\eta,\bar \eta}(\xi)}{ d\xi} \frac{d\eta}{2\pi i \ y(\eta)}F(\eta)
 - \oint_{{\cal C}_{\cal D}^{(\eta)}}\frac{B(p_1,\eta)}{2\pi i \ y(\eta)}F(\eta)
 + \oint_{{\cal C}_{\cal D}^{(\eta)}}\frac{B(p_1,\bar\eta)}{2\pi i \ y(\eta)}F(\eta)
\eeq
where the first integral experiences only
simple poles at the branching points. The residue is unchanged by using l'H\^opital rule,
replacing $B(\xi,p_1)/dy(\xi)$ by $dE_{\xi,\bar\xi}(p_1)/2y(\xi)$, that is,
we have
$$
= - \oint_{{\cal C}_{\cal D}^{(\eta)}} \sum_\alpha \mathop{{\rm Res}}_{\xi\to \mu_\alpha}
\frac{dE_{\xi,\bar\xi}(p_1)\,dy(\xi)\,dE_{\eta,\bar \eta}(\xi)}{ 2 y(\xi)\,d\xi}
\frac{d\eta}{2\pi i \ y(\eta)}F(\eta)
- \oint_{{\cal C}_{\cal D}^{(\eta)}}\frac{B(p_1,\eta)-B(p_1,\bar \eta)}{2\pi i \ y(\eta)}F(\eta)
$$
Pushing contour of integration for $\xi$  around the branch points back through the contour for $\eta$, we pick residues at $\xi=\eta$ and $\xi=\bar\eta$,
which both give the same contribution, and we obtain
\bea
&& - \oint_{{\cal C}_{\cal D}^{(\xi)}>{\cal C}_{\cal D}^{(\eta)}} \frac{dE_{\xi,\bar\xi}(p_1)\,dy(\xi)\,dE_{\eta,\bar \eta}(\xi)}{ 2\pi i \ 2 y(\xi)\,d\xi}
\frac{d\eta}{2\pi i \ y(\eta)}F(\eta) \cr
&& +  \oint_{{\cal C}_{\cal D}^{(\eta)}}  \frac{dE_{\eta,\bar\eta}(p_1)\,dy(\eta)}{  y(\eta)\,d\eta} \frac{d\eta}{2\pi i \ y(\eta)}F(\eta)
 - \oint_{{\cal C}_{\cal D}^{(\eta)}}\frac{B(p_1,\eta)-B(p_1,\bar\eta)}{2\pi i \ y(\eta)}F(\eta)
\eea
We evaluate the last two terms by parts, so it remains only
\beq
 - \oint_{{\cal C}_{\cal D}^{(\xi)}>{\cal C}_{\cal D}^{(\eta)}} \frac{dE_{\xi,\bar\xi}(p_1)\,dy(\xi)\,dE_{\eta,\bar \eta}(\xi)}{ 2\pi i \ 2 y(\xi)\,d\xi}
 \frac{d\eta}{2\pi i \ y(\eta)}F(\eta)
 +  \oint_{{\cal C}_{\cal D}^{(\eta)}}  \frac{dE_{\eta,\bar\eta}(p_1)\,d\eta}{2\pi i \ y(\eta)} \,\frac{\d F(\eta)}{\d\eta}
\eeq

\medskip
We have thus found that
\bea
&& dp_1\,{\d\over \d_{p_1}}\left(\oint_{{\cal C}_{\cal D}^{(\eta)}}\frac{dE_{\eta,\bar\eta}(p_1) d\eta}{2\pi i \ y(\eta)\, dp_1}F(\eta)\right) \cr
&=& - \oint_{{\cal C}_{\cal D}^{(\xi)}>{\cal C}_{\cal D}^{(\eta)}} \frac{dE_{\xi,\bar\xi}(p_1)\,dy(\xi)\,dE_{\eta,\bar \eta}(\xi)}{ 2\pi i \ 2 y(\xi)\,d\xi}
\frac{d\eta}{2\pi i \ y(\eta)}F(\eta)
 +  \oint_{{\cal C}_{\cal D}^{(\eta)}}  \frac{dE_{\eta,\bar\eta}(p_1)\,d\eta}{2\pi i \ y(\eta)} \,\frac{\d F(\eta)}{\d\eta},
\eea
and we can graphically present the action of the derivative as follows:
\be
\begin{picture}(190,55)(50,10)
\thicklines
\put(18,40){\makebox(0,0)[cc]{$\d_{p_1}$}}
\put(25,40){\circle*{2}}
\put(25,40){\vector(1,0){15}}
\put(38,44){\makebox(0,0)[rb]{$\eta$}}
\put(40,40){\circle*{4}}
\put(54,40){\makebox(0,0)[cc]{$\Bigl\{F(\eta)\Bigr\}$}}
\put(73,40){\makebox(0,0)[cc]{$=$}}
\end{picture}
\begin{picture}(0,0)(167,10)
\thicklines
\put(15,40){\makebox(0,0)[cc]{$p_1$}}
\put(20,40){\vector(1,0){12}}
\put(20,40){\circle*{2}}
\put(33,40){\circle*{4}}
\put(34,40){\vector(1,0){12}}
\put(33,38){\makebox(0,0)[ct]{$\xi$}}
\put(47,40){\circle*{4}}
\put(45,38){\makebox(0,0)[rt]{$\eta$}}
\put(61,40){\makebox(0,0)[cc]{$\Bigl\{F(\eta)\Bigr\}$}}
\curvedashes[0.5mm]{2,2}
\curve(20,50, 24,50, 28,49, 30,48, 31,46,33,40)
\put(33,40){\vector(1,-2){0}}
\put(45,38){\makebox(0,0)[rt]{$\eta$}}
\put(35,52){\makebox(0,0)[cb]{$y'(\xi)$}}
\put(77,40){\makebox(0,0)[cc]{$+$}}
\end{picture}
\begin{picture}(0,0)(98,10)
\thicklines
\put(20,40){\makebox(0,0)[cc]{$p_1$}}
\put(25,40){\circle*{2}}
\put(25,40){\vector(1,0){15}}
\put(38,44){\makebox(0,0)[rb]{$\eta$}}
\put(40,40){\circle*{4}}
\put(58,40){\makebox(0,0)[cc]{$\Bigl\{\frac{\d}{\d \eta} F(\eta)\Bigr\}$}}
\end{picture}
\label{Y.5}
\ee

We therefore see that using relation (\ref{Y.5}) we can push the differentiation along the arrowed
edges of a graph. It remains to determine the action of the derivative on internal nonarrowed edges.
But for these edges (since it has two ends), having the term with derivative from the one side, we necessarily
come also to the term with the derivative from the other side; combining these terms, we obtain
$$
\d_p B(p,q)+\d_q B(p,q)=\oint_{{\cal C}_p\cup{\cal C}_q}\frac{B(p,\xi)B(\xi,q)}{2\pi i\ d\xi},
$$
and we can deform this contour to the sum of contours only around the branching points (sum of residues). Then
we can again introduce $y'(\xi)d\xi/y(\xi)$ and integrate out one of the Bergmann kernels (the one that is adjacent
to the point $q$ if $p>q$ or $p$ if $q>p$; recall that, by condition, the points $p$ and $q$ must be comparable).

That is, we have
\be
\begin{picture}(0,55)(180,10)
\thicklines
\put(120,40){\makebox(0,0)[cc]{$(\d_p+\d_q)$}}
\put(140,40){\makebox(0,0)[cc]{$p$}}
\put(145,40){\line(1,0){15}}
\put(145,40){\circle*{2}}
\put(160,40){\circle*{2}}
\put(165,40){\makebox(0,0)[cc]{$q$}}
\put(175,40){\makebox(0,0)[cc]{$=$}}
\put(185,40){\makebox(0,0)[cc]{$p$}}
\put(190,40){\circle*{2}}
\put(190,40){\vector(1,0){10}}
\put(200,40){\circle*{4}}
\put(200,40){\line(1,0){10}}
\put(210,40){\circle*{2}}
\put(215,40){\makebox(0,0)[cc]{$q$}}
\curvedashes[0.5mm]{2,2}
\curve(200,40, 199.5,45, 197,48, 193,51)
\put(200,40){\vector(0,-1){0}}
\put(200,36){\makebox(0,0)[ct]{$r$}}
\put(225,40){\makebox(0,0)[cc]{$\equiv$}}
\put(235,40){\makebox(0,0)[cc]{$p$}}
\put(240,40){\line(1,0){10}}
\put(240,40){\circle*{2}}
\put(250,40){\circle*{4}}
\put(260,40){\vector(-1,0){10}}
\put(260,40){\circle*{2}}
\put(267,40){\makebox(0,0)[cc]{$q\ ,$}}
\curvedashes[0.5mm]{2,2}
\curve(250,40, 249.5,45, 247,48, 243,51)
\put(250,40){\vector(0,-1){0}}
%\put(250,40){\line(0,1){10}}
\put(250,36){\makebox(0,0)[ct]{$r$}}
\end{picture}
\label{variation}
\ee

It becomes clear from the above that we must improve the diagrammatic technique of $\beta$-model in comparison with
the 1MM by including the dashed lines (we do not call them propagators as they have a rather fictitious meaning);
being treated as propagators, they however ensure the proper combinatorics of diagrams.
Indeed, from (\ref{Y.5}) and (\ref{Y.6}) it follows that the derivative action on the ``beginning'' of the dashed line
is null, $\d_p dp=0$, and when this derivative acts on the ``end'' of this line, we merely have
\be
\d_q y'(q)=y''(q),
\label{Y.6}
\ee
which we denote symbolically as {\em two} dashed propagators ending at the same vertex. If we continue to act
by derivatives $\d/\d q$, then, obviously,
when $k$ dashed propagators are terminated at the {\em same} vertex, we have the $k$th order
derivative $y^{(k)}(q)$ corresponding to them.
We then already have three variants of incorporating dashed lines into the play.
Recall that the dashed lines, as the solid nonarrowed lines, can connect only comparable vertices (maybe, the same vertex)
and the starting vertex must necessarily precede the terminating vertex (they may coincide).

The first case is where we have two adjacent solid lines and {\em exactly one} outgoing dashed line. Then, no incoming
dashed lines or extra outgoing dashed lines is possible and the {\em both} solid lines must be arrowed: one is pointed inward
and the other outward. We also distinguish this case by labeling the corresponding vertex by the white dot:
\be
\begin{picture}(0,50)(50,20)
\thicklines
\put(15,40){\makebox(0,0)[cc]{$p$}}
\put(20,40){\vector(1,0){11}}
\put(20,40){\circle*{2}}
\put(33,40){\circle{4}}
\put(33,36){\makebox(0,0)[ct]{$q$}}
\put(35,40){\vector(1,0){11}}
\put(46,40){\circle*{2}}
\put(51,40){\makebox(0,0)[cc]{$r$}}
\curvedashes[0.5mm]{2,2}
%\curve(20,50, 24,49, 28,46, 30,43.5, 31,42, 32.5,40.5)
%\put(33,40){\vector(1,-1){0}}
\curve(46,50, 42,49, 38,46, 36,43.5, 35,42)
\put(46,52.5){\vector(1,1){0}}
\put(60,40){\makebox(0,0)[cc]{$\sim$}}
\put(68,40){\makebox(0,0)[lc]{$\displaystyle\oint_{{\cal C}_{\cal D}^{(q)}}\frac{dq}{2\pi i\ y(q)}$.}}
\end{picture}
\label{Y.8}
\ee

The second case is where we have two adjacent solid lines (one of them is necessarily incoming directed line, but the other
can be either directed outward or be nonarrowed internal or external line), no outgoing dashed line, and
$k\ge1$ incoming dashed lines (this means that we must have at least $k$ white vertices preceding this vertex in the total graph).
In this case, we denote the corresponding vertex by a solid dot:
\be
\begin{picture}(0,50)(50,20)
\thicklines
\put(15,40){\makebox(0,0)[cc]{$p$}}
\put(20,40){\vector(1,0){12}}
\put(20,40){\circle*{2}}
\put(33,40){\circle*{4}}
\put(33,36){\makebox(0,0)[ct]{$q$}}
\put(34,40){\line(1,0){8}}
\put(43,40){\makebox(0,0)[lc]{$(\to)$}}
%\put(46,40){\circle*{2}}
%\put(51,40){\makebox(0,0)[cc]{$r$}}
\curvedashes[0.5mm]{2,2}
\curve(20,50, 24,49, 28,46, 30,43.5, 31,42, 32.5,40.5)
\put(33,40){\vector(1,-1){0}}
\curve(46,50, 42,49, 38,46, 36,43.5, 35,42, 33.5,40.5)
\put(33,40){\vector(-1,-1){0}}
\put(27,50){\circle*{1}}
\put(30.5,51){\circle*{1}}
\put(35.5,51){\circle*{1}}
\put(39,50){\circle*{1}}
\put(33,48){\makebox(0,0)[cb]{$\overbrace{\phantom{---}}$}}
\put(33,61){\makebox(0,0)[cc]{$k$}}
\put(60,40){\makebox(0,0)[cc]{$\sim$}}
\put(68,40){\makebox(0,0)[lc]{$\displaystyle\oint_{{\cal C}_{\cal D}^{(q)}}\frac{y^{(k)}(q)dq}{2\pi i\ y(q)}$,}}
\put(140,40){\makebox(0,0)[cc]{$k\ge1.$}}
\end{picture}
\label{Y.7}
\ee

The last case is when we have just one (incoming arrowed) solid line adjacent to the vertex. We then have {\em exactly one}
outgoing dashed line and $k+1$ incoming dashed lines ($k\ge0$). In this case, we also denote the vertex by the solid dot:
\be
\begin{picture}(0,50)(50,20)
\thicklines
\put(15,40){\makebox(0,0)[cc]{$p$}}
\put(20,40){\vector(1,0){12}}
\put(20,40){\circle*{2}}
\put(33,40){\circle*{4}}
\put(41,40){\makebox(0,0)[cc]{$q$}}
%\put(34,40){\line(1,0){12}}
%\put(46,40){\circle*{2}}
%\put(51,40){\makebox(0,0)[cc]{$r$}}
\curvedashes[0.5mm]{2,2}
\curve(20,50, 24,49, 28,46, 30,43.5, 31,42, 32.5,40.5)
\put(33,40){\vector(1,-1){0}}
\curve(46,50, 42,49, 38,46, 36,43.5, 35,42, 33.5,40.5)
\put(33,40){\vector(-1,-1){0}}
\put(27,50){\circle*{1}}
\put(30.5,51){\circle*{1}}
\put(35.5,51){\circle*{1}}
\put(39,50){\circle*{1}}
\put(33,48){\makebox(0,0)[cb]{$\overbrace{\phantom{---}}$}}
\put(33,61){\makebox(0,0)[cc]{$k$}}
\curve(32.5,39.5, 31,38, 30,35, 29.5,32, 31,30, 33,29.5, 35,30, 36.5,32, 36,35, 35,38, 33.5,39.5)
\put(33,40){\vector(-1,2){0}}
\put(60,40){\makebox(0,0)[cc]{$\sim$}}
\put(68,40){\makebox(0,0)[lc]{$\displaystyle\oint_{{\cal C}_{\cal D}^{(q)}}\frac{y^{(k+1)}(q)dq}{2\pi i\ y(q)}$,}}
\put(140,40){\makebox(0,0)[cc]{$k\ge0,$}}
\end{picture}
\label{Y.9}
\ee

External lines of $W_{k,l}(p_1,\dots, p_s)$ are either the root vertex $p_1$ or nonarrowed propagators $B(q,p_i)$; no
external dashed lines are possible.

Our general rule for assigning white and black colors to vertices is as follows: if there is no factors $y^{(k)}(q)$
standing by the vertex, it is white; if there are such factors, the vertex is painted black.

Now comes the question what happens if we act by the loop insertion operator $d/dV(p)$ on the elements of the
constructed diagrammatic technique.

\subsection{Acting by the loop insertion operator}\label{ss:loop}

We now extend our diagrammatic technique by incorporating the action of loop insertion
operator (\ref{X.6}) on its elements. The action of the loop insertion
operator on the Bergmann differential and its primitive was presented in~\cite{ChEy}, so we do not describe it here.
We can graphically present the action of $\d/\d V(r)$ as
\be
\begin{picture}(240,55)(10,10)
\thicklines
\put(5,40){\makebox(0,0)[cc]{$\frac{\d}{\d V(r)}$}}
\put(20,40){\makebox(0,0)[cc]{$q$}}
\put(25,40){\vector(1,0){14}}
\put(25,40){\circle*{2}}
\put(41,41){\circle{4}}
\put(41,39){\circle*{4}}
\put(47.5,40){\makebox(0,0)[cc]{$p$}}
\put(55,40){\makebox(0,0)[cc]{$=$}}
\put(65,40){\makebox(0,0)[cc]{$q$}}
\put(70,40){\vector(1,0){8}}
\put(70,40){\circle*{2}}
\put(80,40){\circle{4}}
\put(80,36){\makebox(0,0)[ct]{$\xi$}}
\put(82,40){\vector(1,0){8}}
\put(92,41){\circle{4}}
\put(92,39){\circle*{4}}
\put(98.5,40){\makebox(0,0)[cc]{$p,$}}
\put(80,42){\line(0,1){8}}
\put(80,50){\circle*{2}}
\put(80,55){\makebox(0,0)[cc]{$r$}}
\put(125,40){\makebox(0,0)[cc]{$\frac{\d}{\d V(r)}$}}
\put(140,40){\makebox(0,0)[cc]{$q$}}
\put(145,40){\line(1,0){15}}
\put(145,40){\circle*{2}}
\put(160,40){\circle*{2}}
\put(165,40){\makebox(0,0)[cc]{$p$}}
\put(175,40){\makebox(0,0)[cc]{$=$}}
\put(185,40){\makebox(0,0)[cc]{$q$}}
\put(190,40){\vector(1,0){8}}
\put(190,40){\circle*{2}}
\put(200,40){\circle{4}}
\put(200,36){\makebox(0,0)[ct]{$\xi$}}
\put(202,40){\line(1,0){8}}
\put(210,40){\circle*{2}}
\put(215,40){\makebox(0,0)[cc]{$p$}}
\put(200,42){\line(0,1){8}}
\put(200,50){\circle*{2}}
\put(200,55){\makebox(0,0)[cc]{$r$}}
\put(225,40){\makebox(0,0)[cc]{$\equiv$}}
\put(235,40){\makebox(0,0)[cc]{$q$}}
\put(240,40){\line(1,0){8}}
\put(240,40){\circle*{2}}
\put(250,40){\circle{4}}
\put(250,36){\makebox(0,0)[ct]{$\xi$}}
\put(260,40){\vector(-1,0){8}}
\put(260,40){\circle*{2}}
\put(268,40){\makebox(0,0)[cc]{$p\ ,$}}
\put(250,42){\line(0,1){8}}
\put(250,50){\circle*{2}}
\put(250,55){\makebox(0,0)[cc]{$r$}}
\end{picture}
\label{variation1}
\ee
where in the first case we must also take into account the variation of the $y(p)$ factor in the denominator of the
measure of integration in $p$ at the right vertex (irrespectively, white or black), and
the proper contour ordering is assumed. All the appearing vertices are of white color as they do not contain
additional factors of type $y^{(k)}(\xi)$.
In the second case, it is our choice on which of edges to set the arrow. Recall, however, that the points $P$ and $Q$ were already
ordered, as prescribed by the diagram technique. That is, if ``$P> Q$'', we must choose the first variant and if ``$Q> P$'',
we must choose the second variant of arrows arrangement in order to preserve this prescription.

We now calculate the action of $\d/\d V(r)$ on the dashed propagator. Obviously,
\be
\label{dydV}
\frac{\d}{\d V(r)}y^{(k)}(q)=\frac{\d^k}{\d q^k}B(r,q),
\ee
but any attempt to simplify this expression or to reduce it to a combination of
previously introduced diagrammatic elements fails. This means that we must consider it a {\em new} element
of the diagrammatic technique. With a little abuse of notation, we visualize it by preserving
$k$ dashed arrows still landed at the vertex ``$q$" with the added nonarrowed solid line
(third or second,
depending on whether this vertex was of the second (\ref{Y.7}) or third (\ref{Y.9}) kind)
corresponding to the propagator $B(r,q)$.
The vertex then change the coloring from black to white because it contains $y^{(k)}(q)$ factors no more.
We then represent graphically these vertices as
\be
\begin{picture}(0,50)(110,20)
\thicklines
\put(15,40){\makebox(0,0)[cc]{$p$}}
\put(20,40){\vector(1,0){12}}
\put(20,40){\circle*{2}}
\put(33,40){\circle*{4}}
\put(33,36){\makebox(0,0)[ct]{$q$}}
\put(35,40){\line(1,0){8}}
\put(43,40){\makebox(0,0)[lc]{$(\to)$}}
%\put(46,40){\circle*{2}}
\put(18,55){\makebox(0,0)[rc]{$\scriptstyle \frac{\d}{\d V(r)}$}}
\curvedashes[0.5mm]{2,2}
\curve(20,50, 24,49, 28,46, 30,43.5, 31,42, 32.5,40.5)
\put(33,40){\vector(1,-1){0}}
\curve(46,50, 42,49, 38,46, 36,43.5, 35,42, 33.5,40.5)
\put(33,40){\vector(-1,-1){0}}
\put(27,50){\circle*{1}}
\put(30.5,51){\circle*{1}}
\put(35.5,51){\circle*{1}}
\put(39,50){\circle*{1}}
\put(33,48){\makebox(0,0)[cb]{$\overbrace{\phantom{---}}$}}
\put(33,61){\makebox(0,0)[cc]{$k$}}
\put(65,40){\makebox(0,0)[cc]{$=$}}
\end{picture}
\begin{picture}(0,50)(50,20)
\thicklines
\put(15,40){\makebox(0,0)[cc]{$p$}}
\put(20,40){\vector(1,0){11}}
\put(20,40){\circle*{2}}
\put(33,40){\circle{4}}
\put(34,36){\makebox(0,0)[rt]{$q$}}
\put(35,40){\line(1,0){7}}
\put(31.5,38.5){\line(-1,-1){10}}
\put(21.5,28.5){\circle*{2}}
\put(16,29){\makebox(0,0)[lc]{$r$}}
\put(43,40){\makebox(0,0)[lc]{$(\to)$}}
%\put(46,40){\circle*{2}}
%\put(18,55){\makebox(0,0)[cc]{$\scriptstyle \frac{\d}{\d V(r)}$}}
\curvedashes[0.5mm]{2,2}
\curve(20,50, 24,49, 28,46, 30,43.5, 31,42, 32,41)
\put(32,41){\vector(1,-1){0}}
\curve(46,50, 42,49, 38,46, 36,43.5, 35,42)
\put(34,41){\vector(-1,-1){0}}
\put(27,50){\circle*{1}}
\put(30.5,51){\circle*{1}}
\put(35.5,51){\circle*{1}}
\put(39,50){\circle*{1}}
\put(33,48){\makebox(0,0)[cb]{$\overbrace{\phantom{---}}$}}
\put(33,61){\makebox(0,0)[cc]{$k$}}
\put(60,40){\makebox(0,0)[cc]{$\sim$}}
\put(68,40){\makebox(0,0)[lc]{$\displaystyle\oint_{{\cal C}_{\cal D}^{(q)}}\frac{dq}{2\pi i\ y(q)}\frac{\d^k}{\d q^k}B(r,q)$,}}
\put(165,40){\makebox(0,0)[cc]{$k\ge1.$}}
\end{picture}
\label{Y.7*1}
\ee
and
\be
\begin{picture}(0,50)(110,20)
\thicklines
\put(15,40){\makebox(0,0)[cc]{$p$}}
\put(20,40){\vector(1,0){12}}
\put(20,40){\circle*{2}}
\put(33,40){\circle*{4}}
\put(41,40){\makebox(0,0)[cc]{$q$}}
%\put(34,40){\line(1,0){12}}
%\put(46,40){\circle*{2}}
%\put(51,40){\makebox(0,0)[cc]{$r$}}
\put(18,55){\makebox(0,0)[rc]{$\scriptstyle \frac{\d}{\d V(r)}$}}
\curvedashes[0.5mm]{2,2}
\curve(20,50, 24,49, 28,46, 30,43.5, 31,42, 32.5,40.5)
\put(33,40){\vector(1,-1){0}}
\curve(46,50, 42,49, 38,46, 36,43.5, 35,42, 33.5,40.5)
\put(33,40){\vector(-1,-1){0}}
\put(27,50){\circle*{1}}
\put(30.5,51){\circle*{1}}
\put(35.5,51){\circle*{1}}
\put(39,50){\circle*{1}}
\put(33,48){\makebox(0,0)[cb]{$\overbrace{\phantom{---}}$}}
\put(33,61){\makebox(0,0)[cc]{$k$}}
\curve(32.5,39.5, 31,38, 30,35, 29.5,32, 31,30, 33,29.5, 35,30, 36.5,32, 36,35, 35,38, 33.5,39.5)
\put(33,40){\vector(-1,2){0}}
\put(60,40){\makebox(0,0)[cc]{$=$}}
\end{picture}
\begin{picture}(0,50)(50,20)
\thicklines
\put(15,40){\makebox(0,0)[cc]{$p$}}
\put(20,40){\vector(1,0){11}}
\put(20,40){\circle*{2}}
\put(33,40){\circle{4}}
%\put(35,40){\line(1,0){12}}
%\put(47,40){\circle*{2}}
\put(41,38){\makebox(0,0)[ct]{$q$}}
%\put(49,40){\makebox(0,0)[lc]{$r$}}
\put(31.5,38.5){\line(-1,-1){10}}
\put(21.5,28.5){\circle*{2}}
\put(16,29){\makebox(0,0)[lc]{$r$}}
%\put(34,40){\line(1,0){12}}
%\put(46,40){\circle*{2}}
%\put(51,40){\makebox(0,0)[cc]{$r$}}
%\put(18,55){\makebox(0,0)[rc]{$\scriptstyle \frac{\d}{\d V(r)}$}}
\curvedashes[0.5mm]{2,2}
\curve(20,50, 24,49, 28,46, 30,43.5, 31,42, 32,41)
\put(32,41){\vector(1,-1){0}}
\curve(46,50, 42,49, 38,46, 36,43.5, 35,42)
\put(34,41){\vector(-1,-1){0}}
\put(27,50){\circle*{1}}
\put(30.5,51){\circle*{1}}
\put(35.5,51){\circle*{1}}
\put(39,50){\circle*{1}}
\put(33,48){\makebox(0,0)[cb]{$\overbrace{\phantom{---}}$}}
\put(33,61){\makebox(0,0)[cc]{$k$}}
\curve(32,39, 31,38, 30,35, 29.5,32, 31,30, 33,29.5, 35,30, 36.5,32, 36,35, 35,38, 34,39)
\put(33.7,38.6){\vector(-1,2){0}}
\put(60,40){\makebox(0,0)[cc]{$\sim$}}
\put(68,40){\makebox(0,0)[lc]{$\displaystyle\oint_{{\cal C}_{\cal D}^{(q)}}\frac{dq}{2\pi i\ y(q)}\frac{\d^{k+1}}{\d q^{k+1}}B(r,q)$,}}
\put(170,40){\makebox(0,0)[cc]{$k\ge0,$}}
\end{picture}
\label{Y.9*1}
\ee
for the respective cases (\ref{Y.7}) and (\ref{Y.9}). Recall that here necessarily $r>q$.

The reason why we prefer to keep arrowed propagators entering ``white" vertices will be clear after the next step
when we consider the subsequent application of the spatial derivative $\d/\d q$ to the new object $\frac{\d^{k}}{\d q^{k}}B(r,q)$.

Recall that the very application of this operator assumes that we have somewhere in the preceding vertices a vertex of type
(\ref{Y.8}), i.e., the vertex with a new white vertex. Then, successively applying derivative when moving up from the root
over branches of the tree subgraph, we have two possibilities. The first case is where
the vertex ``$r$" is an external vertex of the graph. Then,
when we reach the vertex ``$q$" we just set the extra derivative
$$
\frac{\d}{\d q}\frac{\d^{k}}{\d q^{k}}B(r,q)=\frac{\d^{k+1}}{\d q^{k+1}}B(r,q),
$$
on the corresponding Bergmann kernel, which corresponds to adding an extra incoming dashed
line to either the case (\ref{Y.7*1}) or to the case (\ref{Y.9*1}), depending on the type of the ``white" vertex.
The second and most involved case occurs when
the vertex ``$r$" is an internal vertex of the graph. Then, the derivative must successively act on both
its ends, and we want to express the action on the end ``$r$" in terms of the diagram technique constructed
above. We have from (\ref{variation})
\be
\begin{picture}(0,55)(180,10)
\thicklines
\put(130,40){\makebox(0,0)[cc]{$\displaystyle\left(\frac{\d}{\d q}+\frac{\d}{\d r}\right)\frac{\d^{k}}{\d q^{k}}B(r,q)$}}
\put(175,40){\makebox(0,0)[cc]{$=$}}
\put(185,40){\makebox(0,0)[cc]{$r$}}
\put(190,40){\line(1,0){8}}
\put(190,40){\circle*{2}}
\put(200,40){\circle*{4}}
\put(210,40){\vector(-1,0){8}}
\put(212,40){\circle{4}}
\put(212,36){\makebox(0,0)[ct]{$q$}}
\curvedashes[0.5mm]{2,2}
\curve(200,40, 199.5,45, 197,48, 193,51)
\put(200,40){\vector(0,-1){0}}
%\put(250,40){\line(0,1){10}}
\put(200,36){\makebox(0,0)[ct]{$\xi$}}
\put(217,40){\makebox(0,0)[lc]{$\frac{\d^{k}}{\d q^{k}}$}}
\thinlines
\put(227,48){\vector(-1,0){8}}
\end{picture}
\label{Y.10}
\ee
and it remains to act by $k$ derivatives $\d/\d q$ using the rules formulated in (\ref{Y.5}). We then produce a number of
diagrams, but all of them will be of one of the type indicated above. It becomes clear why we prefer to preserve
incoming dashed lines in notation (\ref{Y.7*1}) and (\ref{Y.9*1}). When applying derivatives $\d/\d q$ in the r.h.s.
of (\ref{Y.10}) a part of these dashed lines that does not act on the last nonarrowed propagator $B(r,\xi)$ appear
again as the derivatives $y^{(i)}(\chi_j)$ at intermediate vertices ``$\chi_j$" $(\xi<\chi_j<q<r)$.

The very last step in constructing the diagram technique is to consider the action of the loop insertion operator
on $\frac{\d^{k}}{\d q^{k}}B(q,r)$. Then, applying now relation (\ref{variation1}), we obtain
\be
\begin{picture}(0,55)(180,10)
\thicklines
\put(130,40){\makebox(0,0)[cc]{$\displaystyle\frac{\d}{\d V(p)}\frac{\d^{k}}{\d q^{k}}B(r,q)$}}
\put(175,40){\makebox(0,0)[cc]{$=$}}
\put(185,40){\makebox(0,0)[cc]{$r$}}
\put(190,40){\line(1,0){8}}
\put(190,40){\circle*{2}}
\put(200,40){\circle{4}}
\put(210,40){\vector(-1,0){8}}
\put(212,40){\circle{4}}
\put(212,36){\makebox(0,0)[ct]{$q$}}
\put(200,42){\line(0,1){8}}
\put(200,50){\circle*{2}}
\put(203,50){\makebox(0,0)[lc]{$p$}}
%\put(250,40){\line(0,1){10}}
\put(200,36){\makebox(0,0)[ct]{$\xi$}}
\put(217,40){\makebox(0,0)[lc]{$\frac{\d^{k}}{\d q^{k}}$}}
\thinlines
\put(227,48){\vector(-1,0){8}}
\end{picture}
\label{Y.11}
\ee
and we observe the appearance of the last remaining structure of the diagram technique---the (``white") vertex
``$\xi$" at which a number $s\le k$ of dashed lines terminate and which have {\em two} incident nonarrowed solid lines
corresponding to the propagators $B(\xi,p)$ and $B(\xi,r)$. This vertex generalizes vertex (\ref{Y.7*1}) in a sense
that the action of derivatives must be now distributed among these {\em two} nonarrowed lines (see (\ref{Y.7*1*1}) below).

It may appear the situation where we have the closed solid loop (the propagator $B(\xi,\bar\xi)$ as shown below)
\be
\begin{picture}(0,50)(100,20)
\thicklines
\put(15,40){\makebox(0,0)[cc]{$q$}}
\put(20,40){\vector(1,0){11}}
\put(20,40){\circle*{2}}
\put(33,40){\circle{4}}
\put(38,40){\makebox(0,0)[lc]{$\xi$}}
%\put(35,40){\line(1,0){9}}
%\put(44,40){\circle*{2}}
%\put(33,38){\line(0,-1){10}}
%\put(33,28){\circle*{2}}
%\put(35,28){\makebox(0,0)[lc]{$r$}}
%\put(47,40){\makebox(0,0)[lc]{$p$}}
%\put(46,40){\circle*{2}}
%\put(18,55){\makebox(0,0)[cc]{$\scriptstyle \frac{\d}{\d V(r)}$}}
\curve(32,39, 31,38, 30,35, 29.5,32, 31,30, 33,29.5, 35,30, 36.5,32, 36,35, 34.5,38.5)
\curvedashes[0.5mm]{2,2}
\curve(20,50, 24,49, 28,46, 30,43.5, 31,42, 32,41)
\put(32,41){\vector(1,-1){0}}
\curve(46,50, 42,49, 38,46, 36,43.5, 35,42)
\put(34,41){\vector(-1,-1){0}}
\put(27,50){\circle*{1}}
\put(30.5,51){\circle*{1}}
\put(35.5,51){\circle*{1}}
\put(39,50){\circle*{1}}
\put(33,48){\makebox(0,0)[cb]{$\overbrace{\phantom{---}}$}}
\put(33,61){\makebox(0,0)[cc]{$k$}}
\put(60,40){\makebox(0,0)[cc]{$\sim$}}
\put(68,40){\makebox(0,0)[lc]{$\displaystyle\oint_{{\cal C}_{\cal D}^{(\xi)}}dE_{\xi,\bar\xi}(q)
\frac{d\xi}{2\pi i\ y(\xi)}\frac{\d^k}{\d \xi^k}B(\xi,\bar\xi)$,}}
\put(200,40){\makebox(0,0)[cc]{$k\ge0.$}}
\end{picture}
\label{Y.12}
\ee
In this case, however, as soon as $k>0$, the derivatives act on {\em both} ends of the propagator $B(\xi,\bar\xi)$,
and we can use (\ref{Y.10}) to distribute them. Therefore, such a diagram (with closed loop of $B$-propagator)
enters the diagram technique only in the case $k=0$.

We are now ready to present the complete diagram technique for multiresolvents of the $\beta$-model.

\section{Feynman diagram rules}\label{s:rules}

We therefore have the following components of the diagrammatic technique.
We absorb all the factors related to Bergmann kernels $B(p,q)$ and $dE_{q,\bar q}(p)$ and all the factors with $y(\xi)$ and
its derivatives into vertices in accordance with the rule: we associate to a vertex the solid arrowed propagator that terminates
at this vertex (recall there is exactly one such propagator) and associate nonarrowed propagator $B(p,q)$ or its
derivatives with the vertex that correspond to the minimal variable among $p$ and $q$. The ordering of vertices is implied
from left to right (as will be assumed in most of appearances below).

{\em The vertices with three adjacent solid lines}
\be
\begin{picture}(0,50)(100,20)
\thicklines
\put(15,40){\makebox(0,0)[cc]{$q$}}
\put(20,40){\vector(1,0){11}}
\put(20,40){\circle*{2}}
\put(33,40){\circle{4}}
\put(34,37){\makebox(0,0)[lt]{$\xi$}}
\put(27,28){\line(1,2){5}}
\put(27,28){\circle*{2}}
\put(20,33){\line(2,1){12}}
\put(20,33){\circle*{2}}
\put(26,27){\makebox(0,0)[rt]{$r$}}
\put(18,32){\makebox(0,0)[rc]{$p$}}
%\put(46,40){\circle*{2}}
%\put(18,55){\makebox(0,0)[cc]{$\scriptstyle \frac{\d}{\d V(r)}$}}
\curvedashes[0.5mm]{2,2}
\curve(20,50, 24,49, 28,46, 30,43.5, 31,42, 32,41)
\put(32,41){\vector(1,-1){0}}
\curve(46,50, 42,49, 38,46, 36,43.5, 35,42)
\put(34,41){\vector(-1,-1){0}}
\put(27,50){\circle*{1}}
\put(30.5,51){\circle*{1}}
\put(35.5,51){\circle*{1}}
\put(39,50){\circle*{1}}
\put(33,48){\makebox(0,0)[cb]{$\overbrace{\phantom{---}}$}}
\put(33,61){\makebox(0,0)[cc]{$k$}}
\put(60,40){\makebox(0,0)[cc]{$\sim$}}
\put(68,40){\makebox(0,0)[lc]{$\displaystyle\oint_{{\cal C}_{\cal D}^{(\xi)}}dE_{\xi,\bar \xi}(q)
\frac{d\xi}{2\pi i\ y(\xi)}\frac{\d^k}{\d \xi^k}\Bigl(B(r,\xi)B(p,\xi)\Bigr)$,}}
\put(220,40){\makebox(0,0)[cc]{$k\ge0.$}}
\end{picture}
\label{Y.7*1*1}
\ee
Here, by construction, $\xi<r$ and $\xi<p$, or $r$ and/or $p$ can be external vertices. If $q$ here is an external vertex,
then $k=0$ and $r$ and $p$ must be also external vertices;
\be
\begin{picture}(0,50)(100,20)
\thicklines
\put(15,40){\makebox(0,0)[cc]{$q$}}
\put(20,40){\vector(1,0){11}}
\put(20,40){\circle*{2}}
\put(33,40){\circle{4}}
\put(37,40){\makebox(0,0)[lc]{$\xi$}}
\curve(32,39, 31,38, 30,35, 29.5,32, 31,30, 33,29.5, 35,30, 36.5,32, 36,35, 34.5,38.5)
\put(60,40){\makebox(0,0)[cc]{$\sim$}}
\put(68,40){\makebox(0,0)[lc]{$\displaystyle\oint_{{\cal C}_{\cal D}^{(\xi)}}dE_{\xi,\bar \xi}(q)
\frac{d\xi}{2\pi i\ y(\xi)}
%\frac{\d^k}{\d \xi^k}
B(\xi,\bar\xi)$,}}
\put(200,40){\makebox(0,0)[cc]{$k\ge0.$}}
\end{picture}
\label{Y.7*1*11}
\ee
The vertex $q$ can be external.
\be
\begin{picture}(0,50)(100,20)
\thicklines
\put(15,40){\makebox(0,0)[cc]{$q$}}
\put(20,40){\vector(1,0){11}}
\put(20,40){\circle*{2}}
\put(33,40){\circle{4}}
\put(34,37){\makebox(0,0)[lt]{$\xi$}}
\put(35,40){\vector(1,0){9}}
\put(44,40){\circle*{2}}
\put(31.5,38.5){\line(-1,-1){10}}
\put(21.5,28.5){\circle*{2}}
\put(16,29){\makebox(0,0)[lc]{$r$}}
\put(47,40){\makebox(0,0)[lc]{$p$}}
%\put(46,40){\circle*{2}}
%\put(18,55){\makebox(0,0)[cc]{$\scriptstyle \frac{\d}{\d V(r)}$}}
\curvedashes[0.5mm]{2,2}
\curve(20,50, 24,49, 28,46, 30,43.5, 31,42, 32,41)
\put(32,41){\vector(1,-1){0}}
\curve(46,50, 42,49, 38,46, 36,43.5, 35,42)
\put(34,41){\vector(-1,-1){0}}
\put(27,50){\circle*{1}}
\put(30.5,51){\circle*{1}}
\put(35.5,51){\circle*{1}}
\put(39,50){\circle*{1}}
\put(33,48){\makebox(0,0)[cb]{$\overbrace{\phantom{---}}$}}
\put(33,61){\makebox(0,0)[cc]{$k$}}
\put(60,40){\makebox(0,0)[cc]{$\sim$}}
\put(68,40){\makebox(0,0)[lc]{$\displaystyle\oint_{{\cal C}_{\cal D}^{(\xi)}}dE_{\xi,\bar\xi}(q)
\frac{d\xi}{2\pi i\ y(\xi)}\frac{\d^k}{\d \xi^k}B(r,\xi)$,}}
\put(210,40){\makebox(0,0)[cc]{$k\ge0,\quad r>\xi.$}}
\end{picture}
\label{Y.7*1*2}
\ee
Here $r$ can be external vertex. If $q$ is an external vertex, then $k=0$ and $r$ is also an external vertex.
\be
\begin{picture}(0,50)(100,20)
\thicklines
\put(15,40){\makebox(0,0)[cc]{$q$}}
\put(20,40){\vector(1,0){11}}
\put(20,40){\circle*{2}}
\put(33,40){\circle{4}}
\put(32,37){\makebox(0,0)[rt]{$\xi$}}
\put(35,40){\vector(1,0){9}}
\put(44,40){\circle*{2}}
\put(34.5,38.5){\line(1,-1){10}}
\put(44.5,28.5){\circle*{2}}
\put(46,29){\makebox(0,0)[lc]{$r$}}
\put(47,40){\makebox(0,0)[lc]{$p$}}
\put(60,40){\makebox(0,0)[cc]{$\sim$}}
\put(68,40){\makebox(0,0)[lc]{$\displaystyle\oint_{{\cal C}_{\cal D}^{(\xi)}}dE_{\xi,\bar\xi}(q)\frac{d\xi}{2\pi i\ y(\xi)}$.}}
\put(180,40){\makebox(0,0)[cc]{$\xi>r.$}}
%\put(180,40){\makebox(0,0)[cc]{$k\ge0.$}}
\end{picture}
\label{Y.7*1*3}
\ee
Here $q$ can be an external vertex (and $r$ is not by condition).
\be
\begin{picture}(0,50)(100,20)
\thicklines
\put(15,40){\makebox(0,0)[cc]{$q$}}
\put(20,40){\vector(1,0){11}}
\put(20,40){\circle*{2}}
\put(33,40){\circle{4}}
\put(32,37){\makebox(0,0)[rt]{$\xi$}}
\put(35,40){\vector(1,0){9}}
\put(44,40){\circle*{2}}
\put(34.5,38.5){\vector(1,-1){10}}
\put(44.5,28.5){\circle*{2}}
\put(46,29){\makebox(0,0)[lc]{$r$}}
\put(47,40){\makebox(0,0)[lc]{$p$}}
%\put(46,40){\circle*{2}}
%\put(18,55){\makebox(0,0)[cc]{$\scriptstyle \frac{\d}{\d V(r)}$}}
%\curvedashes[0.5mm]{2,2}
%\curve(20,50, 24,49, 28,46, 30,43.5, 31,42, 32,41)
%\put(32,41){\vector(1,-1){0}}
%\curve(46,50, 42,49, 38,46, 36,43.5, 35,42)
%\put(34,41){\vector(-1,-1){0}}
%\put(27,50){\circle*{1}}
%\put(30.5,51){\circle*{1}}
%\put(35.5,51){\circle*{1}}
%\put(39,50){\circle*{1}}
%\put(33,48){\makebox(0,0)[cb]{$\overbrace{\phantom{---}}$}}
%\put(33,61){\makebox(0,0)[cc]{$k$}}
\put(60,40){\makebox(0,0)[cc]{$\sim$}}
\put(68,40){\makebox(0,0)[lc]{$\displaystyle\oint_{{\cal C}_{\cal D}^{(\xi)}}dE_{\xi,\bar\xi}(q)\frac{d\xi}{2\pi i\ y(\xi)}$.}}
%\put(180,40){\makebox(0,0)[cc]{$k\ge0.$}}
\end{picture}
\label{Y.7*1*4}
\ee
Here $q$ can be an external vertex.

{\em The vertices with two adjacent solid lines}
\be
\begin{picture}(0,50)(100,20)
\thicklines
\put(15,40){\makebox(0,0)[cc]{$p$}}
\put(20,40){\vector(1,0){12}}
\put(20,40){\circle*{2}}
\put(33,40){\circle*{4}}
\put(34,36){\makebox(0,0)[ct]{$q$}}
\put(31.5,38.5){\line(-1,-1){10}}
\put(21.5,28.5){\circle*{2}}
\put(16,29){\makebox(0,0)[lc]{$r$}}
\curvedashes[0.5mm]{2,2}
\curve(20,50, 24,49, 28,46, 30,43.5, 31,42, 32.5,40.5)
\put(33,40){\vector(1,-1){0}}
\curve(46,50, 42,49, 38,46, 36,43.5, 35,42, 33.5,40.5)
\put(33,40){\vector(-1,-1){0}}
\put(27,50){\circle*{1}}
\put(30.5,51){\circle*{1}}
\put(35.5,51){\circle*{1}}
\put(39,50){\circle*{1}}
\put(33,48){\makebox(0,0)[cb]{$\overbrace{\phantom{---}}$}}
\put(33,61){\makebox(0,0)[cc]{$k$}}
\put(60,40){\makebox(0,0)[cc]{$\sim$}}
\put(68,40){\makebox(0,0)[lc]{$\displaystyle\oint_{{\cal C}_{\cal D}^{(q)}}dE_{q,\bar q}(p)\frac{y^{(k)}(q)dq}{2\pi i\ y(q)}B(r,q)$,}}
\put(180,40){\makebox(0,0)[cc]{$k\ge1.$}}
\end{picture}
\label{Y.7*2*1}
\ee
Here by construction $q<r$ and $r$ can be an external vertex, while $p$ cannot be an external vertex.
\be
\begin{picture}(0,50)(100,20)
\thicklines
\put(15,40){\makebox(0,0)[cc]{$p$}}
\put(20,40){\vector(1,0){12}}
\put(20,40){\circle*{2}}
\put(33,40){\circle*{4}}
\put(33,36){\makebox(0,0)[ct]{$q$}}
\put(34,40){\vector(1,0){12}}
\put(46,40){\circle*{2}}
%\put(43,40){\makebox(0,0)[lc]{$(\to)$}}
%\put(46,40){\circle*{2}}
\put(46,36){\makebox(0,0)[cc]{$r$}}
\curvedashes[0.5mm]{2,2}
\curve(20,50, 24,49, 28,46, 30,43.5, 31,42, 32.5,40.5)
\put(33,40){\vector(1,-1){0}}
\curve(46,50, 42,49, 38,46, 36,43.5, 35,42, 33.5,40.5)
\put(33,40){\vector(-1,-1){0}}
\put(27,50){\circle*{1}}
\put(30.5,51){\circle*{1}}
\put(35.5,51){\circle*{1}}
\put(39,50){\circle*{1}}
\put(33,48){\makebox(0,0)[cb]{$\overbrace{\phantom{---}}$}}
\put(33,61){\makebox(0,0)[cc]{$k$}}
\put(60,40){\makebox(0,0)[cc]{$\sim$}}
\put(68,40){\makebox(0,0)[lc]{$\displaystyle\oint_{{\cal C}_{\cal D}^{(q)}}dE_{q,\bar q}(p)\frac{y^{(k)}(q)dq}{2\pi i\ y(q)}$.}}
\put(170,40){\makebox(0,0)[cc]{$k\ge1.$}}
\end{picture}
\label{Y.7*2*2}
\ee
Here $p$ cannot be an external vertex.
\be
\begin{picture}(0,50)(100,20)
\thicklines
\put(15,40){\makebox(0,0)[cc]{$p$}}
\put(20,40){\vector(1,0){11}}
\put(20,40){\circle*{2}}
\put(33,40){\circle{4}}
\put(41,40){\makebox(0,0)[cc]{$q$}}
%\put(34,40){\line(1,0){12}}
%\put(46,40){\circle*{2}}
%\put(51,40){\makebox(0,0)[cc]{$r$}}
%\put(18,55){\makebox(0,0)[rc]{$\scriptstyle \frac{\d}{\d V(r)}$}}
\put(31.5,38.5){\line(-1,-1){10}}
\put(21.5,28.5){\circle*{2}}
\put(16,29){\makebox(0,0)[lc]{$r$}}
\curvedashes[0.5mm]{2,2}
\curve(20,50, 24,49, 28,46, 30,43.5, 31,42, 32,41)
\put(32,41){\vector(1,-1){0}}
\curve(46,50, 42,49, 38,46, 36,43.5, 35,42)
\put(34,41){\vector(-1,-1){0}}
\put(27,50){\circle*{1}}
\put(30.5,51){\circle*{1}}
\put(35.5,51){\circle*{1}}
\put(39,50){\circle*{1}}
\put(33,48){\makebox(0,0)[cb]{$\overbrace{\phantom{---}}$}}
\put(33,61){\makebox(0,0)[cc]{$k$}}
\curve(32,39, 31,38, 30,35, 29.5,32, 31,30, 33,29.5, 35,30, 36.5,32, 36,35, 35,38, 34,39)
\put(33.7,38.6){\vector(-1,2){0}}
\put(60,40){\makebox(0,0)[cc]{$\sim$}}
\put(68,40){\makebox(0,0)[lc]{$\displaystyle\gamma\oint_{{\cal C}_{\cal D}^{(q)}}dE_{q,\bar q}(p)
\frac{dq}{2\pi i\ y(q)}\frac{\d^{k+1}}{\d q^{k+1}}B(r,q)$,}}
\put(220,40){\makebox(0,0)[cc]{$k\ge0.$}}
\end{picture}
\label{Y.7*2*3}
\ee
Here always $q<r$ or $r$ is an external vertex. If $p$ is an external vertex, then $k=0$ and $r$ is also an
external vertex.
\be
\begin{picture}(0,50)(100,20)
\thicklines
\put(15,40){\makebox(0,0)[cc]{$p$}}
\put(20,40){\vector(1,0){11}}
\put(20,40){\circle*{2}}
\put(33,40){\circle{4}}
\put(33,36){\makebox(0,0)[ct]{$q$}}
\put(35,40){\vector(1,0){11}}
\put(46,40){\circle*{2}}
\put(51,40){\makebox(0,0)[cc]{$r$}}
\curvedashes[0.5mm]{2,2}
%\curve(20,50, 24,49, 28,46, 30,43.5, 31,42, 32.5,40.5)
%\put(33,40){\vector(1,-1){0}}
\curve(46,50, 42,49, 38,46, 36,43.5, 35,42)
\put(46,52.5){\vector(1,1){0}}
\put(60,40){\makebox(0,0)[cc]{$\sim$}}
\put(68,40){\makebox(0,0)[lc]{$\displaystyle\gamma\oint_{{\cal C}_{\cal D}^{(q)}}dE_{q,\bar q}(p)\frac{dq}{2\pi i\ y(q)}$.}}
\end{picture}
\label{Y.7*2*4}
\ee
Here $p$ can be an external vertex.

{\em The vertex with one adjacent solid line}
\be
\begin{picture}(0,50)(100,20)
\thicklines
\put(15,40){\makebox(0,0)[cc]{$p$}}
\put(20,40){\vector(1,0){12}}
\put(20,40){\circle*{2}}
\put(33,40){\circle*{4}}
\put(41,40){\makebox(0,0)[cc]{$q$}}
%\put(34,40){\line(1,0){12}}
%\put(46,40){\circle*{2}}
%\put(51,40){\makebox(0,0)[cc]{$r$}}
\curvedashes[0.5mm]{2,2}
\curve(20,50, 24,49, 28,46, 30,43.5, 31,42, 32.5,40.5)
\put(33,40){\vector(1,-1){0}}
\curve(46,50, 42,49, 38,46, 36,43.5, 35,42, 33.5,40.5)
\put(33,40){\vector(-1,-1){0}}
\put(27,50){\circle*{1}}
\put(30.5,51){\circle*{1}}
\put(35.5,51){\circle*{1}}
\put(39,50){\circle*{1}}
\put(33,48){\makebox(0,0)[cb]{$\overbrace{\phantom{---}}$}}
\put(33,61){\makebox(0,0)[cc]{$k$}}
\curve(32.5,39.5, 31,38, 30,35, 29.5,32, 31,30, 33,29.5, 35,30, 36.5,32, 36,35, 35,38, 33.5,39.5)
\put(33,40){\vector(-1,2){0}}
\put(60,40){\makebox(0,0)[cc]{$\sim$}}
\put(68,40){\makebox(0,0)[lc]{$\displaystyle\gamma\oint_{{\cal C}_{\cal D}^{(q)}}dE_{q,\bar q}(p)\frac{y^{(k+1)}(q)dq}{2\pi i\ y(q)}$,}}
\put(180,40){\makebox(0,0)[cc]{$k\ge0.$}}
\end{picture}
\label{Y.7*3*1}
\ee
If $p$ here is external, then $k=0$.

When calculating $W_g(p_1,\dots,p_s)$ we take the sum over {\em all possible} graphs with one external leg $dE_{q,\bar q}(p_1)$ and all
other external legs to be $B(\xi_j,p_i)$ and, possibly, their derivatives w.r.t. internal variables $\xi_j$ in accordance with the above
rules such that arrowed propagators constitute a maximum directed tree subgraph with the root at $p_1$ and the bold nonarrowed lines can connect
only comparable vertices (and may have derivatives only at their inner endpoints) and arrowed dashed lines can also connect only
comparable vertices (the arrow is then directed along the arrows on the tree subgraph). The diagrams enter the sum with the standard
symmetry coefficients.

In full analogy with the 1MM case, the order of integration contours is prescribed by the order of vertices in the subtree: the closer is the
vertex to the root, the more outer is the integration contour. In contrast to the 1MM case, the integration cannot be reduced to taking residues at
the branching points only; all internal integrations can be nevertheless reduced to sums of residues, but these sums may now include
residues at zeros of the additional polynomial $M(p)$ on the nonphysical sheet and, possibly, at the point $\infty_-$.

In complete analogy with the 1MM case,
in the next section, we use the $H$-operator introduced in~\cite{ChEy} in order to invert the action of the loop insertion operator and
obtain the expression for the free energy itself.

\section{Inverting the loop insertion operator. Free energy  \label{s:freeBeta}}

\subsection{The $H$-operator \label{ss:H}}

We now use the operator that is in a sense inverse to loop insertion operator (\ref{X.6}) and was introduced
in~\cite{ChEy} in the 1MM case. It has the form\footnote{This definition works well
when acting on 1-forms regular at infinities. Otherwise (say, in the case of
$W_0(p)$), the integral in the third term must be regularized, e.g., by replacing it by the contour integral around the logarithmic
cut stretched between two infinities.}
\be
H\cdot=\frac12\res_{\infty_+}V(x)\cdot\ -\frac12\res_{\infty_-}V(x)\cdot\
-t_0\int_{\infty_-}^{\infty_+}\cdot\ -\sum_{i=1}^{n-1}S_i\oint_{B_i}\cdot.
\label{H}
\ee
The arrangement of the integration contours see in Fig.~1. We now calculate the action of $H$ on the Bergmann
bidifferential $B(x,q)$ using again the Riemann bilinear identities.
We first note that $B(x,q)=\left(\d_xdE_{x,q_0}(q)\right)dx$ and we can evaluate residues at infinities
by parts. Then, since $dE_{x,q_0}(q)$ is regular at infinities, for $V'(x)$ we substitute $2y(x)+2t_0/x$ as $x\to\infty_+$
and $-2y(x)+2t_0/x$ as $x\to\infty_-$ thus obtaining
\bea
&&-\res_{\infty_+}\left(y(x)+\frac{t_0}{x}\right)dE_{x,q_0}(q)dx+\res_{\infty_-}\left(-y(x)+\frac{t_0}{x}\right)dE_{x,q_0}(q)dx
\nonumber
\\
&&\qquad \Bigl.-t_0dE_{x,q_0}(q)\Bigr|_{x=\infty_-}^{x=\infty_+}-\sum_{i=1}^{n-1}S_i\oint_{B_i}B(q,x),
\label{Z.1}
\eea
whence the cancelation of terms containing $t_0$ is obvious, and it remains to take the combination of residues
at infinities involving $y(x)$. For this, we cut the surface along $A$- and $B$-cycles taking into account the residue at
$x=q$. The boundary integrals on two sides of the cut at $B_i$ then differ by $dE_{x,q_0}(q)-dE_{x+\oint_{A_i},q_0}(q)=0$,
while the integrals on two sides of the cut at $A_i$ differ by  $dE_{x,q_0}(q)-dE_{x+\oint_{B_i},q_0}(q)=\oint_{B_i}B(q,x)$,
and we obtain for the boundary term the expression
$$
\sum_{i=1}^{n-1}\oint_{A_i}y(x)dx\oint_{B_i}B(q,\xi),
$$
which exactly cancel the last term in (\ref{Z.1}). It remains only the contribution from the pole at $x=q$, which is just
$-y(q)$. We have therefore proved that
\be
H\cdot B(\cdot,q)=-y(q)dq.
\label{HB}
\ee

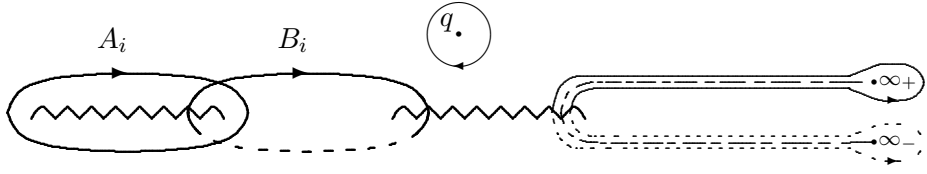
\begin{figure}[tb]
%\hspace*{2cm}
%\epsfysize=6cm
\vskip .2in
\setlength{\unitlength}{0.8mm}%
\begin{picture}(0,40)(15,15)
\thinlines
\put(135,43){\circle{10}}
\put(135,43){\circle*{1}}
\put(133,45){\makebox(0,0)[cc]{$q$}}
\put(134,37.7){\vector(-1,0){0}}
\thicklines
\curve(60,30, 62,33.5, 65,35, 70,36, 80,36.5, 90,36, 95,35, 98,33.5, 100,30, 98,26.5, 95,25, 90,24, 80,23.5, 70,24, 65,25, 62,26.5, 60,30)
\put(80,36.5){\vector(1,0){0}}
\curve(64,29, 66,31, 68,29, 70,31, 72,29, 74,31, 76,29, 78,31, 80,29, 82,31, 84,29, 86,31, 88,29, 90,31, 92,29, 94,31, 96,29)
\put(80,40){\makebox(0,0)[rb]{$A_i$}}
%\curve(120,30, 122,33.5, 125,35, 130,36, 140,36.5, 150,36, 155,35, 158,33.5, 160,30, 158,26.5, 155,25, 150,24, 140,23.5, 130,24, 125,25, 122,26.5, 120,30)
\curve(124,29, 126,31, 128,29, 130,31, 132,29, 134,31, 136,29, 138,31, 140,29, 142,31, 144,29, 146,31, 148,29, 150,31, 152,29, 154,31, 156,29)
\put(110,40){\makebox(0,0)[rb]{$B_i$}}
\curve(92,26.5, 90,30, 92,33.5, 95,35, 100,36, 110,36.5, 120,36, 125,35, 128,33.5, 130,30, 128,26.5)
\put(110,36.5){\vector(1,0){0}}
{\curvedashes[1mm]{0,1,2}
\curve(92,33.5, 90,30, 92,26.5, 95,25, 100,24, 110,23.5, 120,24, 125,25, 128,26.5, 130,30, 128,33.5)}
\put(204,35){\circle*{1}}
\put(204,25){\circle*{1}}
{\thinlines
\curvedashes[1mm]{1,3,2}
\curve(204,35, 157,35, 154,34, 153,33, 152,30, 153,27, 154,26, 157,25, 204,25)
\curvedashes[1mm]{5,1}
\curve(204,35, 157,35, 154,34, 153,33, 152,30, 153,27, 154,26, 157,25, 204,25)}
\thinlines
%\put(180,37){\makebox(0,0)[cb]{$C_L$}}
%\put(207,40){\makebox(0,0)[cb]{$C_\Lambda$}}
\put(208,35){\makebox(0,0)[cc]{${\scriptstyle \infty_+}$}}
\put(208,25){\makebox(0,0)[cc]{${\scriptstyle \infty_-}$}}
\put(208,32){\vector(1,0){0}}
\put(208,22){\vector(1,0){0}}
\curve(153.5,30, 154,31.8, 154.5,32.8, 155.5,33.5, 157.5,34, 164,34, 199,34, 200,34, 201,33.5, 205,32, 209,32, 210,32.3, 211,32.8, 212,34, 212.3,35)
\curve(150.5,30, 151,32.2, 151.5,33.7, 152.5,35, 154.5,36, 162,36, 199,36, 200,36, 201,36.5, 205,38, 209,38, 210,37.7, 211,37.2, 212,36, 212.3,35)
\curvedashes[0.5mm]{2,1}
\curve(153.5,30, 154,28.2, 154.5,27.2, 155.5,26.5, 157.5,26, 164,26, 199,26, 200,26, 201,26.5, 205,28, 209,28, 210,27.7, 211,27.2, 212,26, 212.3,25)
\curve(150.5,30, 151,27.8, 151.5,26.3, 152.5,25, 154.5,24, 162,24, 199,24, 200,24, 201,23.5, 205,22, 209,22, 210,22.3, 211,22.8, 212,24, 212.3,25)
\end{picture}
%\centerline{\epsfxsize=11.5cm\epsffile{wdvv11.eps}}
\caption{The arrangement of integration contours on the Riemann surface.}
\label{fi:cuts}
\end{figure}

Let us now consider the action of $H$ on $W_{k,l}(\cdot)$ subsequently evaluating the action of loop insertion operator (\ref{X.6}) on the
result. Note first that the only result of action of $\d/\d V(p)$ on the operator $H$ itself are derivatives
$\d V(x)/\d V(p)=-1/(p-x)$ (and recall that by definition $|p|>|x|$, i.e., instead of evaluating residues at infinities one should
take residues at $x=p$, and we obtain
\be
\frac{\d}{\d V(p)}\left(H\cdot W_{k,l}(\cdot)\right)=W_{k,l}(p)+H\cdot W_{k,l}(\cdot,p).
\label{Z.2}
\ee
For the second term, due to the symmetry of $W_{k,l}(p,q)$, we may choose the point $p$ to be the root of the tree subgraphs. Then,
the operator $H$ always acts on $B(\cdot,\xi)$ (or, possibly, on its derivatives w.r.t. $\xi$)
where $\xi$ are integration variables of internal vertices.

Let us recall the action of $\d/\d V(q)$ on the elements of the Feynman diagram technique in Sec.~\ref{s:rules}.
Here we have three different cases.
\begin{itemize}
    \item When acting on the arrowed propagator followed by a (white or black) vertex, we use the first relation  in (\ref{variation1}).
    \item When acting on nonarrowed internal propagator $\frac{\d^k}{\d q^k}B(p,q)$, $p\ge q$,
$k\ge 0$, we apply relation (\ref{Y.11}) {\em without} subsequent representing the action of the derivative $\frac{\d^k}{\d q^k}$
as a sum of diagrams. We have no external $B$-lines as we act on the one-loop resolvent.
    \item Eventually, when acting on dashed lines coming to a black vertex, using relation (\ref{dydV})
    we obtain expression in (\ref{Y.7*1}); the action on dashed lines coming to a white vertex is null.
\end{itemize}

We now consider the {\em inverse} action of the $H$-operator in all three cases.

In the first case where it exists an outgoing arrowed propagator $dE_{p,q_0}(\xi)$ (we can have only
one such arrowed propagator as one line is external), then we can push the integration contour for $\xi$ through the one for $p$;
the only contribution comes from the pole at $\xi=p$ (with the {\em opposite} sign due to the choice of contour directions in
Fig.~1. We then obtain the following graphical representation for the action of the operator $H$ in the first case:
\be
\begin{picture}(240,30)(10,30)
\thicklines
\put(65,40){\makebox(0,0)[cc]{$q$}}
\put(70,40){\vector(1,0){8}}
\put(70,40){\circle*{2}}
\put(80,40){\circle{4}}
\put(80,36){\makebox(0,0)[ct]{$\xi$}}
\put(82,40){\vector(1,0){8}}
\put(92,41){\circle{4}}
\put(92,39){\circle*{4}}
\put(98.5,40){\makebox(0,0)[cc]{$p$}}
\put(80,42){\line(0,1){8}}
\put(80,50){\circle*{2}}
\put(80,55){\makebox(0,0)[cc]{$H\cdot$}}
\put(110,40){\makebox(0,0)[cc]{$=$}}
\put(120,40){\makebox(0,0)[cc]{$-$}}
\put(130,40){\makebox(0,0)[cc]{$q$}}
\put(135,40){\vector(1,0){14}}
\put(135,40){\circle*{2}}
\put(151,41){\circle{4}}
\put(151,39){\circle*{4}}
\put(157.5,40){\makebox(0,0)[cc]{$p$}}
\end{picture}
\label{chopping}
\ee

In the second case, the vertex $\xi$ in (\ref{Y.11})
is an innermost vertex (i.e., there is no arrowed edges coming out of it). The 1-form $y(\xi)d\xi$ arising under the action of $H$
(\ref{HB}) cancels the corresponding form in the integration expression, and the residue vanishes being nonsingular at the
branching point. Graphically, we have
\be
\begin{picture}(0,55)(150,10)
\thicklines
\put(85,40){\makebox(0,0)[cc]{$r$}}
\put(90,40){\line(1,0){8}}
\put(90,40){\circle*{2}}
\put(100,40){\circle{4}}
\put(110,40){\vector(-1,0){8}}
\put(112,40){\circle{4}}
\put(112,36){\makebox(0,0)[ct]{$q$}}
\put(100,42){\line(0,1){8}}
\put(100,50){\circle*{2}}
\put(103,50){\makebox(0,0)[lc]{$H\cdot$}}
%\put(250,40){\line(0,1){10}}
\put(100,36){\makebox(0,0)[ct]{$\xi$}}
\put(117,40){\makebox(0,0)[lc]{$\frac{\d^{k}}{\d q^{k}}$}}
\thinlines
\put(127,48){\vector(-1,0){8}}
\put(135,40){\makebox(0,0)[lc]{$=\ 0,$}}
\put(165,40){\makebox(0,0)[lc]{$k\ge0.$}}
\end{picture}
\label{chopping2}
\ee

Eventually, in the third case, the inversion is rather easy to produce. Indeed, action of $H$-operator just erases the new $B$-propagator
arising in (\ref{Y.7*1}) simultaneously changing back the color of the vertex from white to black:
\be
\begin{picture}(0,50)(80,20)
\thicklines
\put(15,40){\makebox(0,0)[cc]{$p$}}
\put(20,40){\vector(1,0){11}}
\put(20,40){\circle*{2}}
\put(33,40){\circle{4}}
\put(32,36){\makebox(0,0)[rt]{$q$}}
\put(35,40){\line(1,0){7}}
\put(33,38){\line(0,-1){10}}
\put(33,28){\circle*{2}}
\put(35,28){\makebox(0,0)[lc]{$H\cdot$}}
\put(43,40){\makebox(0,0)[lc]{$(\to)$}}
%\put(46,40){\circle*{2}}
%\put(18,55){\makebox(0,0)[cc]{$\scriptstyle \frac{\d}{\d V(r)}$}}
\curvedashes[0.5mm]{2,2}
\curve(20,50, 24,49, 28,46, 30,43.5, 31,42, 32,41)
\put(32,41){\vector(1,-1){0}}
\curve(46,50, 42,49, 38,46, 36,43.5, 35,42)
\put(34,41){\vector(-1,-1){0}}
\put(27,50){\circle*{1}}
\put(30.5,51){\circle*{1}}
\put(35.5,51){\circle*{1}}
\put(39,50){\circle*{1}}
\put(33,48){\makebox(0,0)[cb]{$\overbrace{\phantom{---}}$}}
\put(33,61){\makebox(0,0)[cc]{$k$}}
\put(65,40){\makebox(0,0)[cc]{$=$}}
\end{picture}
\begin{picture}(0,50)(20,20)
\thicklines
\put(15,40){\makebox(0,0)[cc]{$p$}}
\put(20,40){\vector(1,0){12}}
\put(20,40){\circle*{2}}
\put(33,40){\circle*{4}}
\put(33,36){\makebox(0,0)[ct]{$q$}}
\put(34,40){\line(1,0){8}}
\put(43,40){\makebox(0,0)[lc]{$(\to)$}}
%\put(46,40){\circle*{2}}
%\put(18,55){\makebox(0,0)[rc]{$\scriptstyle \frac{\d}{\d V(r)}$}}
\curvedashes[0.5mm]{2,2}
\curve(20,50, 24,49, 28,46, 30,43.5, 31,42, 32.5,40.5)
\put(33,40){\vector(1,-1){0}}
\curve(46,50, 42,49, 38,46, 36,43.5, 35,42, 33.5,40.5)
\put(33,40){\vector(-1,-1){0}}
\put(27,50){\circle*{1}}
\put(30.5,51){\circle*{1}}
\put(35.5,51){\circle*{1}}
\put(39,50){\circle*{1}}
\put(33,48){\makebox(0,0)[cb]{$\overbrace{\phantom{---}}$}}
\put(33,61){\makebox(0,0)[cc]{$k$}}
\put(65,40){\makebox(0,0)[lc]{$k\ge1$.}}
\end{picture}
\label{chopping3}
\ee

For $H_q\cdot W_k(q,p)=H_q\cdot \frac{\d}{\d V(q)}W_k(p)$, we obtain that for each arrowed edge, on which the action of
(\ref{X.6}) produces the new (white) vertex, the inverse action of $H_q\cdot$ gives the factor $-1$,
on each nonarrowed edge,
on which the action of (\ref{X.6}) produces the new vertex accordingly to (\ref{Y.11}),
the inverse action of $H_q\cdot$ just gives zero, and at each black vertex, at which the action of (\ref{X.6})
changes the color to white and adds a new $B$-propagator, the inverse action of $H_q\cdot$ gives the factor $+1$.

As the total number
of arrowed edges coincides with the total number of vertices and the contributions of black vertices are opposite to the
contributions of arrowed edges, the total factor on which the diagram is multiplied is exactly minus the number of white
vertices, which is $2k+l-1$ for {\em any} graph contributing to $W_{k,l}(p)$. We then have
$$
H_q\cdot W_{k,l}(q,p)=-(2k+l-1)W_k(p)
$$
and, combining with (\ref{Z.2}), we just obtain
\be
\frac{\d}{\d V(p)}\left(H_q\cdot W_{k,l}(q)\right)=-(2-2k-l)\frac{\d}{\d V(p)}{\cal F}_{k,l},
\label{Z.3}
\ee
and, since all the dependence on filling fractions and $t_0$ is fixed by condition (\ref{Z.5}),
we conclude that
\be
{\cal F}_{k,l}=\frac{1}{2k+l-2}H\cdot W_{k,l}.
\label{fin}
\ee
This is our final answer for the free energy. It permits us to calculate {\em all}
${\cal F}_{k,l}$ except the contribution at $k=1,l=0$ (torus approximation
in the 1MM) and the second-order correction in $\gamma$ (the term ${\cal F}_{0,2}$).
The term ${\cal F}_{1,0}$ was calculated by a direct integration in~\cite{Chekh}. We devote the special section below for the calculation of
the term ${\cal F}_{0,2}$.
{\em All other} orders can be consistently calculated. For this,
we only introduce the {\em new vertex} $\circ$\hskip-0.85ex$\cdot$
at which we place the (nonlocal) integral term
$\oint_{{\cal C}^{(\xi)}}\frac{d\xi}{2\pi i}\frac{\int_{\bar\xi}^\xi y(s)ds}{y(\xi)}$.
In the 1MM case, although this term was also nonlocal, it was possible to shift the starting point at the branching
point $\mu_\alpha$ in the vicinity of every branching point; here it is no more the case and we must consider global
integrations. Note, however, that it is {\em only} for the very last integration for which we must introduce nonlocal terms;
all internal integrations can be performed by taking residues at branching points and at the zeros of the polynomial $M(p)$.
The examples of diagrams are collected in the next section.

\subsection{Low-order corrections}\label{ss:low}

We begin with presenting several low-order corrections to the free energy.
\be
\hbox{
\begin{picture}(240,20)(10,30)
\thicklines
\put(35,40){\makebox(0,0)[cc]{$-1\cdot {\cal F}_{0,1}$}}
\put(55,40){\makebox(0,0)[cc]{$=$}}
\put(65,40){\circle*{1.5}}
\put(65,40){\circle{3}}
{\curvedashes[0.5mm]{2,2}
\curve(65.4,41.2, 66.4,43.4, 68.4,44.6, 70,45, 71.6,44.6, 73.6,43.4, 74.6,41.2, 74.8,40,
 74.6,38.8, 73.6,36.6, 71.6,35.4, 70,35, 68.4,35.4, 66.4,36.6, 65.4,38.8)
}
\put(74,38){\vector(0,-1){0}}
\end{picture}
}
\label{F01}
\ee
This gives
$$
{\cal F}_{0,1}=-\oint_{{\cal C}_{\cal D}^{(q)}}\frac{\int^q y}{y(q)}y'(q)\frac{dq}{2\pi i}=
\oint_{{\cal C}_{\cal D}^{(q)}}y(q)\log y(q)\frac{dq}{2\pi i},
$$
i.e., we obtain the semiclassical Dyson term~\cite{Dyson}.

Next two diagrams cannot be presented in the free-energy form, so we present the expressions for
the one-loop resolvents:
\be
\hbox{
\begin{picture}(240,20)(10,30)
\thicklines
\put(25,40){\makebox(0,0)[cc]{$W_{1,0}(p)$}}
\put(40,40){\makebox(0,0)[cc]{$=$}}
\put(55,40){\circle*{1.5}}
\put(53,40){\makebox(0,0)[cr]{$p$}}
\put(55,40){\vector(1,0){8.5}}
\put(65,40){\circle{3}}
%{\curvedashes[0.5mm]{2,2}
\curve(65.4,41.2, 66.4,43.4, 68.4,44.6, 70,45, 71.6,44.6, 73.6,43.4, 74.6,41.2, 74.8,40,
 74.6,38.8, 73.6,36.6, 71.6,35.4, 70,35, 68.4,35.4, 66.4,36.6, 65.4,38.8)
%}
%\put(74,38){\vector(0,-1){0}}
\end{picture}
}
\label{W10}
\ee
\be
\hbox{
\begin{picture}(240,20)(10,30)
\thicklines
\put(5,40){\makebox(0,0)[cc]{$W_{0,2}(p)$}}
\put(20,40){\makebox(0,0)[cc]{$=$}}
%\put(38,40){\makebox(0,0)[cc]{$2$}}
\put(35,40){\vector(1,0){8.5}}
\put(35,40){\circle*{1.5}}
\put(33,40){\makebox(0,0)[cr]{$2\ p$}}
\put(45,40){\circle{3}}
%\put(45,40){\circle*{1.5}}
%\put(45,40){\makebox(0,0)[cc]{$\cdot$}}
{\curvedashes[0.5mm]{2,2}
\curve(45.4,41.2, 46.4,43.4, 48.4,44.6, 50,45, 51.6,44.6, 53.6,43.4, 54.6,41.2)
}
\curve(45.4,38.8, 46.4,36.6, 48.4,35.4, 50,35, 51.6,35.4, 53.6,36.6, 54.6,38.8)
%\put(50,40){\circle{10}}
\put(55,40){\circle*{3}}
\put(52.5,45){\vector(1,0){0}}
\put(52.5,35.5){\vector(1,0){0}}
\put(56.5,40){\vector(1,0){7}}
\put(65,40){\circle*{3}}
{\curvedashes[0.5mm]{2,2}
\curve(65.4,41.2, 66.4,43.4, 68.4,44.6, 70,45, 71.6,44.6, 73.6,43.4, 74.6,41.2, 74.8,40,
 74.6,38.8, 73.6,36.6, 71.6,35.4, 70,35, 68.4,35.4, 66.4,36.6, 65.4,38.8)
}
\put(74,38){\vector(0,-1){0}}
\put(83,40){\makebox(0,0)[cc]{$+$}}
\end{picture}
\begin{picture}(0,20)(190,30)
\thicklines
\put(35,40){\vector(1,0){8.5}}
\put(35,40){\circle*{1.5}}
\put(33,40){\makebox(0,0)[rc]{$2\ p$}}
\put(45,40){\circle{3}}
%\put(45,40){\circle*{1.5}}
%\put(45,40){\makebox(0,0)[cc]{$\cdot$}}
{\curvedashes[0.5mm]{2,2}
\curve(45.4,41.2, 46.4,43.4, 48.4,44.6, 50,45, 51.6,44.6, 53.6,43.4, 54.6,41.2)
}
\curve(45.4,38.8, 46.4,36.6, 48.4,35.4, 50,35, 51.6,35.4, 53.6,36.6, 54.6,38.8)
%\put(50,40){\circle{10}}
\put(55,40){\circle*{3}}
\put(52.5,45){\vector(1,0){0}}
\put(52.5,35.5){\vector(1,0){0}}
%\put(56.5,40){\vector(1,0){7}}
%\put(65,40){\circle*{3}}
{\curvedashes[0.5mm]{2,2}
\curve(55.4,41.2, 56.4,43.4, 58.4,44.6, 60,45, 61.6,44.6, 63.6,43.4, 64.6,41.2, 64.8,40,
 64.6,38.8, 63.6,36.6, 61.6,35.4, 60,35, 58.4,35.4, 56.4,36.6, 55.4,38.8)
}
\put(64,38){\vector(0,-1){0}}
\end{picture}
\begin{picture}(0,20)(240,30)
\thicklines
\put(125,40){\makebox(0,0)[cc]{$+$}}
\put(140,40){\circle*{3}}
\curvedashes[0.5mm]{2,2}
\curve(139.6,41.2, 138.6,43.4, 136.6,44.6, 135,45, 133.4,44.6, 131.4,43.4, 130.4,41.2, 130.2,40,
 130.4,38.8, 131.4,36.6, 133.4,35.4, 135,35, 136.6,35.4, 138.6,36.6, 139.6,38.8)
\put(130,38){\vector(0,-1){0}}
\put(148.5,40){\vector(-1,0){7}}
\put(150,40){\circle{3}}
\put(152,50){\makebox(0,0)[lc]{$p$}}
\put(150,50){\circle*{1.5}}
\put(150,50){\vector(0,-1){8.5}}
\put(151.5,40){\vector(1,0){7}}
\put(160,40){\circle*{3}}
\curve(160.4,41.2, 161.4,43.4,163.4,44.6, 165,45, 166.6,44.6, 168.6,43.4, 169.6,41.2, 169.8,40,
 169.6,38.8, 168.6,36.6, 166.6,35.4, 165,35, 163.4,35.4, 161.4,36.6, 160.4,38.8)
\put(169,38){\vector(0,-1){0}}
\end{picture}
}
\label{W02}
\ee

We also present two first ``regular'' terms of the free-energy expansion:
\be
\hbox{
\begin{picture}(240,50)(10,10)
\thicklines
\put(15,40){\makebox(0,0)[cc]{$1\cdot {\cal F}_{1,1}$}}
\put(30,40){\makebox(0,0)[cc]{$=$}}
%\put(38,40){\makebox(0,0)[cc]{$2$}}
%\put(35,40){\vector(1,0){8.5}}
\put(45,40){\circle*{1.5}}
\put(45,40){\circle{3}}
%\put(45,40){\circle*{1.5}}
%\put(45,40){\makebox(0,0)[cc]{$\cdot$}}
{\curvedashes[0.5mm]{2,2}
\curve(45.4,41.2, 46.4,43.4, 48.4,44.6, 50,45, 51.6,44.6, 53.6,43.4, 54.6,41.2)
}
\curve(45.4,38.8, 46.4,36.6, 48.4,35.4, 50,35, 51.6,35.4, 53.6,36.6, 54.6,38.8)
%\put(50,40){\circle{10}}
\put(55,40){\circle*{3}}
\put(52.5,45){\vector(1,0){0}}
\put(52.5,35.5){\vector(1,0){0}}
\put(56.5,40){\vector(1,0){7}}
\put(65,40){\circle{3}}
%{\curvedashes[0.5mm]{2,2}
\curve(65.4,41.2, 66.4,43.4, 68.4,44.6, 70,45, 71.6,44.6, 73.6,43.4, 74.6,41.2, 74.8,40,
 74.6,38.8, 73.6,36.6, 71.6,35.4, 70,35, 68.4,35.4, 66.4,36.6, 65.4,38.8)
%}
%\put(74,38){\vector(0,-1){0}}
\put(80,40){\makebox(0,0)[cc]{$+$}}
\end{picture}
\begin{picture}(0,20)(210,10)
\thicklines
\put(45,40){\circle*{1.5}}
\put(45,40){\circle{3}}
%\put(45,40){\circle*{1.5}}
%\put(45,40){\makebox(0,0)[cc]{$\cdot$}}
%{\curvedashes[0.5mm]{2,2}
\curve(45.4,41.2, 46.4,43.4, 48.4,44.6, 50,45, 51.6,44.6, 53.6,43.4, 54.6,41.2)
%}
\curve(45.4,38.8, 46.4,36.6, 48.4,35.4, 50,35, 51.6,35.4, 53.6,36.6, 54.6,38.8)
%\put(50,40){\circle{10}}
\put(55,40){\circle{3}}
%\put(52.5,45){\vector(1,0){0}}
\put(52.5,35.5){\vector(1,0){0}}
\put(56.5,40){\vector(1,0){7}}
\put(65,40){\circle*{3}}
{\curvedashes[0.5mm]{2,2}
\curve(65.4,41.2, 66.4,43.4, 68.4,44.6, 70,45, 71.6,44.6, 73.6,43.4, 74.6,41.2, 74.8,40,
 74.6,38.8, 73.6,36.6, 71.6,35.4, 70,35, 68.4,35.4, 66.4,36.6, 65.4,38.8)
}
\put(74,38){\vector(0,-1){0}}
\put(80,40){\makebox(0,0)[cc]{$+$}}
\end{picture}
\begin{picture}(0,20)(170,10)
\thicklines
\put(45,40){\circle*{1.5}}
\put(45,40){\circle{3}}
%\put(45,40){\circle*{1.5}}
%\put(45,40){\makebox(0,0)[cc]{$\cdot$}}
%{\curvedashes[0.5mm]{2,2}
\curve(45.4,41.2, 46.4,43.4, 48.4,44.6, 50,45, 51.6,44.6, 53.6,43.4, 54.6,41.2)
%}
\curve(45.4,38.8, 46.4,36.6, 48.4,35.4, 50,35, 51.6,35.4, 53.6,36.6, 54.6,38.8)
%\put(50,40){\circle{10}}
\put(55,40){\circle{3}}
%\put(52.5,45){\vector(1,0){0}}
\put(52.5,35.5){\vector(1,0){0}}
%\put(56.5,40){\vector(1,0){7}}
%\put(65,40){\circle*{3}}
{\curvedashes[0.5mm]{2,2}
\curve(55.4,41.2, 56.4,43.4, 58.4,44.6, 60,45, 61.6,44.6, 63.6,43.4, 64.6,41.2, 64.8,40,
 64.6,38.8, 63.6,36.6, 61.6,35.4, 60,35, 58.4,35.4, 56.4,36.6, 55.4,38.8)
}
\put(64,38){\vector(0,-1){0}}
\end{picture}
\begin{picture}(0,20)(220,10)
\thicklines
\put(120,40){\makebox(0,0)[cc]{$+2$}}
\put(140,40){\circle*{3}}
{\curvedashes[0.5mm]{2,2}
\curve(139.6,41.2, 138.6,43.4, 136.6,44.6, 135,45, 133.4,44.6, 131.4,43.4, 130.4,41.2, 130.2,40,
 130.4,38.8, 131.4,36.6, 133.4,35.4, 135,35, 136.6,35.4, 138.6,36.6, 139.6,38.8)
 }
\put(130,38){\vector(0,-1){0}}
\put(148.5,40){\vector(-1,0){7}}
\put(150,40){\circle{3}}
\put(150,40){\circle*{1.5}}
\put(151.5,40){\vector(1,0){7}}
\put(160,40){\circle*{3}}
\curve(160.4,41.2, 161.4,43.4,163.4,44.6, 165,45, 166.6,44.6, 168.6,43.4, 169.6,41.2, 169.8,40,
 169.6,38.8, 168.6,36.6, 166.6,35.4, 165,35, 163.4,35.4, 161.4,36.6, 160.4,38.8)
%\put(169,38){\vector(0,-1){0}}
\end{picture}
\begin{picture}(0,20)(240,35)
\thicklines
\put(85,40){\makebox(0,0)[cc]{$+$}}
\put(95,40){\circle{3}}
\put(95,40){\circle*{1.5}}
\put(106.5,40){\oval(20,20)[r]}
%\put(105,50){\circle*{2}}
\put(116,38){\vector(0,-1){0}}
{\curvedashes[0.5mm]{2,2}
\curve(105,48.5, 105.5,40, 105,31.5)
 }
\put(105.2,38){\vector(0,-1){0}}
\put(105,50){\circle{3}}
\put(105,30){\circle*{3}}
\curve(95.2,41.4, 96.8,45, 100,48.2, 103.4,49.8)
\curve(95.2,38.6, 96.8,35, 100,31.8, 103.4,30.2)
\put(100,48.5){\vector(1,1){0}}
\end{picture}
\begin{picture}(0,20)(200,35)
\thicklines
\put(85,40){\makebox(0,0)[cc]{$+$}}
\put(95,40){\circle{3}}
\put(95,40){\circle*{1.5}}
\put(106.5,40){\oval(20,20)[r]}
%\put(105,50){\circle*{2}}
%\put(116,38){\vector(0,-1){0}}
\put(105,48.5){\vector(0,-1){17}}
\put(105,50){\circle{3}}
\put(105,30){\circle*{3}}
\curve(95.2,41.4, 96.8,45, 100,48.2, 103.4,49.8)
{\curvedashes[0.5mm]{2,2}
\curve(95.2,38.6, 96.8,35, 100,31.8, 103.4,30.2)
}
\put(100,48.5){\vector(1,1){0}}
\put(100,31.5){\vector(1,-1){0}}
\end{picture}
}
\label{F11}
\ee

\be
\hbox{
\begin{picture}(240,20)(10,30)
\thicklines
\put(15,40){\makebox(0,0)[cc]{$2\cdot {\cal F}_{2,0}$}}
\put(30,40){\makebox(0,0)[cc]{$=$}}
\put(38,40){\makebox(0,0)[cc]{$2$}}
\put(45,40){\circle{3}}
\put(45,40){\circle*{1.5}}
%\put(45,40){\makebox(0,0)[cc]{$\cdot$}}
\curve(45.4,41.2, 46.4,43.4, 48.4,44.6, 50,45, 51.6,44.6, 53.6,43.4, 54.6,41.2)
\curve(45.4,38.8, 46.4,36.6, 48.4,35.4, 50,35, 51.6,35.4, 53.6,36.6, 54.6,38.8)
%\put(50,40){\circle{10}}
\put(55,40){\circle{3}}
\put(53,45){\vector(1,0){0}}
\put(56.5,40){\vector(1,0){7}}
\put(65,40){\circle{3}}
\curve(65.4,41.2, 66.4,43.4, 68.4,44.6, 70,45, 71.6,44.6, 73.6,43.4, 74.6,41.2, 74.8,40,
 74.6,38.8, 73.6,36.6, 71.6,35.4, 70,35, 68.4,35.4, 66.4,36.6, 65.4,38.8)
%\put(70,40){\circle{10}}
%%
\put(85,40){\makebox(0,0)[cc]{$+2$}}
\put(95,40){\circle{3}}
\put(95,40){\circle*{1.5}}
\put(106.5,40){\oval(20,20)[r]}
%\put(105,50){\circle*{2}}
\put(105,48.5){\vector(0,-1){17}}
\put(105,50){\circle{3}}
\put(105,30){\circle{3}}
\curve(95.2,41.4, 96.8,45, 100,48.2, 103.4,49.8)
\curve(95.2,38.6, 96.8,35, 100,31.8, 103.4,30.2)
\put(100,48.5){\vector(1,1){0}}
%\put(105,30){\circle*{2}}
%%
\put(125,40){\makebox(0,0)[cc]{$+$}}
\put(140,40){\circle{3}}
\curve(139.6,41.2, 138.6,43.4, 136.6,44.6, 135,45, 133.4,44.6, 131.4,43.4, 130.4,41.2, 130.2,40,
 130.4,38.8, 131.4,36.6, 133.4,35.4, 135,35, 136.6,35.4, 138.6,36.6, 139.6,38.8)
\put(148.5,40){\vector(-1,0){7}}
\put(150,40){\circle{3}}
\put(150,40){\circle*{1.5}}
\put(151.5,40){\vector(1,0){7}}
\put(160,40){\circle{3}}
\curve(160.4,41.2, 161.4,43.4,163.4,44.6, 165,45, 166.6,44.6, 168.6,43.4, 169.6,41.2, 169.8,40,
 169.6,38.8, 168.6,36.6, 166.6,35.4, 165,35, 163.4,35.4, 161.4,36.6, 160.4,38.8)
\end{picture}
}
\label{F20}
\ee

The free-energy diagrammatic terms corresponding to resolvents (\ref{W10}) and (\ref{W02})
vanish (for (\ref{W10}) see \cite{ChEy}). The free-energy term ${\cal F}_{1,0}$ is the subleading
correction in 1MM calculated in \cite{Chekh}. It therefore remains only to calculate
(\ref{W02}) subsequently integrating it to obtain the corresponding free-energy contribution ${\cal F}_{0,2}$.

\section{Calculating ${\cal F}_{0,2}$}\label{s:F02}

We begin with demonstrating that the diagrammatical expression for ${\cal F}_{0,2}$ of the form
$$
\hbox{
\begin{picture}(240,20)(10,30)
\thicklines
%\put(5,40){\makebox(0,0)[cc]{$W_{0,2}(p)$}}
%\put(20,40){\makebox(0,0)[cc]{$=$}}
\put(38,40){\makebox(0,0)[cc]{$2$}}
%\put(35,40){\vector(1,0){8.5}}
%\put(35,40){\circle*{1.5}}
%\put(33,40){\makebox(0,0)[cr]{$p$}}
\put(45,40){\circle{3}}
\put(45,40){\circle*{1.5}}
%\put(45,40){\makebox(0,0)[cc]{$\cdot$}}
{\curvedashes[0.5mm]{2,2}
\curve(45.4,41.2, 46.4,43.4, 48.4,44.6, 50,45, 51.6,44.6, 53.6,43.4, 54.6,41.2)
}
\curve(45.4,38.8, 46.4,36.6, 48.4,35.4, 50,35, 51.6,35.4, 53.6,36.6, 54.6,38.8)
%\put(50,40){\circle{10}}
\put(55,40){\circle*{3}}
\put(52.5,45){\vector(1,0){0}}
\put(52.5,35.5){\vector(1,0){0}}
\put(56.5,40){\vector(1,0){7}}
\put(65,40){\circle*{3}}
{\curvedashes[0.5mm]{2,2}
\curve(65.4,41.2, 66.4,43.4, 68.4,44.6, 70,45, 71.6,44.6, 73.6,43.4, 74.6,41.2, 74.8,40,
 74.6,38.8, 73.6,36.6, 71.6,35.4, 70,35, 68.4,35.4, 66.4,36.6, 65.4,38.8)
}
\put(74,38){\vector(0,-1){0}}
\put(85,40){\makebox(0,0)[cc]{$+$}}
\end{picture}
\begin{picture}(0,20)(180,30)
\thicklines
%\put(35,40){\vector(1,0){8.5}}
%\put(35,40){\circle*{1.5}}
%\put(33,40){\makebox(0,0)[rc]{$p$}}
\put(38,40){\makebox(0,0)[cc]{$2$}}
\put(45,40){\circle{3}}
\put(45,40){\circle*{1.5}}
%\put(45,40){\makebox(0,0)[cc]{$\cdot$}}
{\curvedashes[0.5mm]{2,2}
\curve(45.4,41.2, 46.4,43.4, 48.4,44.6, 50,45, 51.6,44.6, 53.6,43.4, 54.6,41.2)
}
\curve(45.4,38.8, 46.4,36.6, 48.4,35.4, 50,35, 51.6,35.4, 53.6,36.6, 54.6,38.8)
%\put(50,40){\circle{10}}
\put(55,40){\circle*{3}}
\put(52.5,45){\vector(1,0){0}}
\put(52.5,35.5){\vector(1,0){0}}
%\put(56.5,40){\vector(1,0){7}}
%\put(65,40){\circle*{3}}
{\curvedashes[0.5mm]{2,2}
\curve(55.4,41.2, 56.4,43.4, 58.4,44.6, 60,45, 61.6,44.6, 63.6,43.4, 64.6,41.2, 64.8,40,
 64.6,38.8, 63.6,36.6, 61.6,35.4, 60,35, 58.4,35.4, 56.4,36.6, 55.4,38.8)
}
\put(64,38){\vector(0,-1){0}}
\end{picture}
\begin{picture}(0,20)(230,30)
\thicklines
\put(120,40){\makebox(0,0)[cc]{$+$}}
\put(140,40){\circle*{3}}
\curvedashes[0.5mm]{2,2}
\curve(139.6,41.2, 138.6,43.4, 136.6,44.6, 135,45, 133.4,44.6, 131.4,43.4, 130.4,41.2, 130.2,40,
 130.4,38.8, 131.4,36.6, 133.4,35.4, 135,35, 136.6,35.4, 138.6,36.6, 139.6,38.8)
\put(130,38){\vector(0,-1){0}}
\put(148.5,40){\vector(-1,0){7}}
\put(150,40){\circle{3}}
\put(150,40){\circle*{1.5}}
%\put(152,50){\makebox(0,0)[lc]{$p$}}
%\put(150,50){\vector(0,-1){8.5}}
\put(151.5,40){\vector(1,0){7}}
\put(160,40){\circle*{3}}
\curve(160.4,41.2, 161.4,43.4,163.4,44.6, 165,45, 166.6,44.6, 168.6,43.4, 169.6,41.2, 169.8,40,
 169.6,38.8, 168.6,36.6, 166.6,35.4, 165,35, 163.4,35.4, 161.4,36.6, 160.4,38.8)
\put(169,38){\vector(0,-1){0}}
\end{picture}
}
%\label{F02}
$$
vanishes. For this, we observe that the first two terms are
$$
2\oint_{{\cal C}_{\cal D}^{(q)}}\oint_{{\cal C}_{\cal D}^{(\xi)}}
\frac{\int_{\bar q}^q y}{y(q)}{\frac{\d}{\d q}}dE_{\xi,\bar \xi}(q)\frac{y'(\xi)d\xi}{2\pi i\ y(\xi)},
$$
and we can integrate by parts to set the derivative w.r.t. $q$ on the exceptional vertex.
To do this, however, we must first interchange the order of contour integrations (as we have
the integral of $y$ inside the outer integration in $q$).
This interchanging yields the additional term
\be
-2\oint_{{\cal C}_{\cal D}^{(\xi)}}\frac{\int_{\bar\xi}^\xi y}{y(\xi)}\frac{\d}{\d \xi}\frac{y'(\xi)d\xi}{2\pi i\ y(\xi)},
\label{F02:1}
\ee
and, upon integrating by parts, we have
$$
\frac{\d}{\d q}\frac{\int_{\bar q}^q y}{y(q)}=2-\frac{y'(q)\int_{\bar q}^q y}{y^2(q)}.
$$
The constant part does not contribute, while for the second part we introduce the notation
$$
\hbox{
\begin{picture}(240,20)(-80,30)
\thicklines
\put(45,40){\makebox(0,0)[cc]{$\frac{y'(q)\int_{\bar q}^q y}{y^2(q)}\equiv$}}
\put(65,40){\circle{3}}
\put(65,40){\circle*{1.5}}
{\curvedashes[0.5mm]{2,2}
\curve(65.4,41.2, 66.4,43.4, 68.4,44.6, 70,45, 71.6,44.6, 73.6,43.4, 74.6,41.2, 74.8,40,
 74.6,38.8, 73.6,36.6, 71.6,35.4, 70,35, 68.4,35.4, 66.4,36.6, 65.4,38.8)
}
\put(74,38){\vector(0,-1){0}}
\end{picture}
}
$$
The remaining contribution can be then graphically depicted as
\be
\hbox{
\begin{picture}(0,20)(130,30)
\thicklines
\put(120,40){\makebox(0,0)[cc]{$2$}}
\put(140,40){\circle*{3}}
\curvedashes[0.5mm]{2,2}
\curve(139.6,41.2, 138.6,43.4, 136.6,44.6, 135,45, 133.4,44.6, 131.4,43.4, 130.4,41.2, 130.2,40,
 130.4,38.8, 131.4,36.6, 133.4,35.4, 135,35, 136.6,35.4, 138.6,36.6, 139.6,38.8)
\put(130,38){\vector(0,-1){0}}
\put(148.5,40){\vector(-1,0){7}}
\put(150,40){\circle{3}}
\put(150,40){\circle*{1.5}}
\curve(150.4,41.2, 151.4,43.4,153.4,44.6, 155,45, 156.6,44.6, 158.6,43.4, 159.6,41.2, 159.8,40,
 159.6,38.8, 158.6,36.6, 156.6,35.4, 155,35, 153.4,35.4, 151.4,36.6, 150.4,38.8)
\put(159,38){\vector(0,-1){0}}
\end{picture}
}
\label{F02:2}
\ee
where we assume now that the order of appearance of vertices (from left to right) implies the contour
ordering for the corresponding integrations.

As concerning the third term, if we collapse the outer integration (the one with the integral in $y$ term) to
the support~${\cal D}$, it gives zero upon integration because it contains no singularities,
and we remain with two terms appearing when passing through two other integration contours.
Both these terms are of the same form as (\ref{F02:2}) except they both come with the factors $-1$ and they differ
by the integration order. Therefore, combining with (\ref{F02:2}), we have the contribution
$$
\hbox{
\begin{picture}(0,20)(180,30)
\thicklines
\put(120,40){\makebox(0,0)[cc]{$-$}}
\put(140,40){\circle{3}}
\put(140,40){\circle*{1.5}}
\curvedashes[0.5mm]{2,2}
\curve(139.6,41.2, 138.6,43.4, 136.6,44.6, 135,45, 133.4,44.6, 131.4,43.4, 130.4,41.2, 130.2,40,
 130.4,38.8, 131.4,36.6, 133.4,35.4, 135,35, 136.6,35.4, 138.6,36.6, 139.6,38.8)
\put(130,38){\vector(0,-1){0}}
\put(141.5,40){\vector(1,0){7}}
\put(150,40){\circle*{3}}
\curve(150.4,41.2, 151.4,43.4,153.4,44.6, 155,45, 156.6,44.6, 158.6,43.4, 159.6,41.2, 159.8,40,
 159.6,38.8, 158.6,36.6, 156.6,35.4, 155,35, 153.4,35.4, 151.4,36.6, 150.4,38.8)
\put(159,38){\vector(0,-1){0}}
\end{picture}
\begin{picture}(0,20)(130,30)
\thicklines
\put(120,40){\makebox(0,0)[cc]{$+$}}
\put(140,40){\circle*{3}}
\curvedashes[0.5mm]{2,2}
\curve(139.6,41.2, 138.6,43.4, 136.6,44.6, 135,45, 133.4,44.6, 131.4,43.4, 130.4,41.2, 130.2,40,
 130.4,38.8, 131.4,36.6, 133.4,35.4, 135,35, 136.6,35.4, 138.6,36.6, 139.6,38.8)
\put(130,38){\vector(0,-1){0}}
\put(148.5,40){\vector(-1,0){7}}
\put(150,40){\circle{3}}
\put(150,40){\circle*{1.5}}
\curve(150.4,41.2, 151.4,43.4,153.4,44.6, 155,45, 156.6,44.6, 158.6,43.4, 159.6,41.2, 159.8,40,
 159.6,38.8, 158.6,36.6, 156.6,35.4, 155,35, 153.4,35.4, 151.4,36.6, 150.4,38.8)
\put(159,38){\vector(0,-1){0}}
\end{picture}
}
$$
which can be evaluated by taking the residue at $q=\xi$ (doubled because of two residues at two sheets).
It gives the integral over ${\cal D}$ with the integrand $2\int y\cdot (y')^2\cdot y^{-3}$
Together with the first residue term (\ref{F02:1}), it gives
$$
2\oint_{{\cal C}_{\cal D}^{(q)}}\frac{dq}{2\pi i}
\frac{\int_{\bar q}^q y}{y(q)}\left(\frac{y'(q)}{y(q)}\cdot\frac{y'(q)}{y(q)}-\frac{\d }{\d q}\left(\frac{y'(q)}{y(q)}\right)\right)=
-2\oint_{{\cal C}_{\cal D}^{(q)}}\frac{dq}{2\pi i}
{\hbox{$\int_{\bar q}^q y$}}\cdot\frac{\d }{\d q}\left(\frac{y'(q)}{y^2(q)}\right)
$$
and, integrating by parts, we just obtain $2\oint_{{\cal C}_{\cal D}^{(q)}}\frac{dq}{2\pi i}\frac{y'(q)}{y^(q)}=2n$,
i.e., the constant independent on the potential.

We can now make a guess for the actual ${\cal F}_{0,2}$. It seems very plausible to expect it to be of the Polyakov's
anomaly form $\iint R\frac1{\Delta} R$, where $R$ is the curvature and $1/\Delta$ is the Green's function for
the Laplace operator, which in our case is the logarithm of the Prime form. The curvature is expressed through
the function $y$ as $R\sim y'/y$. That is, we have two natural candidates for ${\cal F}_{0,2}$:
$$
{\cal F}_{0,2}\sim \iint dq\,dp\frac{y'(q)}{y(q)}\log E(p,q)\frac{y'(p)}{y(p)},
$$
where $E$ is the prime form, or
$$
{\cal F}_{0,2}\sim \iint dq\,dp\log y(q) B(p,q)\log y(p),
$$
but neither of these expressions is well defined. The first one develops the logarithmic cut at $p=q$ and cannot be
written in a contour-independent way; moreover, both these expressions are divergent when integrating along the
support. We therefore must find another representation imitating this term. A good choice is when we integrate by part
only once, we then have the expression of the form
\be
\oint_{{\cal C}_{\cal D}^{(q)}}\frac{dq}{2\pi i}
\frac{y'(q)}{y(q)}\int_{D}dE_{q,\bar q}(p)\log y(p) dp,
\label{F02:3}
\ee
where the second integral is taken along just the eigenvalue support $D$.

Variation of the logarithmic term in (\ref{F02:3}) can be presented already in the form
of the contour integral. In what follows, we also systematically use that
$$
\int_D dq B(p,q)\log y(q)=-\oint_{{\cal C}_{\cal D}^{(q)}}\frac{dq}{2\pi i}dE_{q,\bar q}(p)\frac{y'(q)}{y(q)}
$$
for the point $p$ outside the contour ${\cal C}_{\cal D}^{(q)}$.

Action of $\frac{\d}{\d V(r)}$ on ``outer'' combination $y'(q)/y(q)$ gives $\d_qB(r,q)/y(q)$, which, upon
integration by parts, gives
\bea
&{}&
-\oint_{{\cal C}_{\cal D}^{(q)}}B(r,q)\frac1{y(q)}\oint_{{\cal C}_{\cal D}^{(p)}}B(q,p)\log y(p)
=-\oint_{{\cal C}_{\cal D}^{(q)}}dE_{q,\bar q}(r)\frac{\d}{\d q}\frac1{y(q)}\oint_{{\cal C}_{\cal D}^{(p)}}dE_{p,\bar p}(q)
\frac{y'(p)}{y(p)}
\nonumber
\\
&{}&\quad
=\oint_{{\cal C}_{\cal D}^{(q)}}dE_{q,\bar q}(r)\frac{y'(q)}{y^2(q)}\oint_{{\cal C}_{\cal D}^{(p)}}dE_{p,\bar p}(q)\frac{y'(p)}{y(p)}
-\oint_{{\cal C}_{\cal D}^{(q)}}dE_{q,\bar q}(r)\frac1{y(q)}\frac{\d}{\d q}\oint_{{\cal C}_{\cal D}^{(p)}}dE_{p,\bar p}(q),
\label{F02:4}
\eea
and in the second term we recognize the two first diagrams of (\ref{W02}) (with factors $-1$).

Action of $\frac{\d}{\d V(r)}$ on ``inner'' $\log y(p)$ gives $B(r,p)/y(p)$, and in order
to correspond to our diagrammatic technique we must interchange the order of integration
w.r.t. $q$ and $p$. And for the last action of $\frac{\d}{\d V(r)}$ on $dE_{p,\bar p}(q)$, we use
(\ref{variation1}). That is, implying the contour ordering as above (from left to right) and indicating
explicitly all the $y$-factors in the vertices, we graphically obtain for the last two variations:
$$
\hbox{
\begin{picture}(0,40)(180,30)
\thicklines
\put(120,40){\makebox(0,0)[cc]{$r$}}
\put(125,40){\circle*{1.5}}
\put(125,40){\vector(1,0){25}}
\put(150,40){\vector(-1,1){15}}
\put(150,40){\line(-1,-1){10}}
\put(150,40){\circle*{1.5}}
\put(135,55){\circle*{1.5}}
\put(140,30){\circle*{1.5}}
\put(129,57){\makebox(0,0)[cc]{$\frac{y'}{y}$}}
\put(138,60){\makebox(0,0)[cc]{$q$}}
\put(130,28){\makebox(0,0)[cc]{\small ${\log y}$}}
\put(143,25){\makebox(0,0)[cc]{$p$}}
\put(155,45){\makebox(0,0)[cc]{$\frac 1y$}}
\put(153,35){\makebox(0,0)[cc]{$s$}}
\end{picture}
\begin{picture}(0,40)(120,30)
\thicklines
\put(110,40){\makebox(0,0)[cc]{$+$}}
\put(120,40){\makebox(0,0)[cc]{$r$}}
\put(125,40){\circle*{1.5}}
\put(125,40){\line(1,0){25}}
\put(150,40){\vector(-1,1){15}}
%\put(150,40){\line(-1,-1){10}}
\put(150,40){\circle*{1.5}}
\put(135,55){\circle*{1.5}}
%\put(140,30){\circle*{1.5}}
\put(129,57){\makebox(0,0)[cc]{$\frac{y'}{y}$}}
\put(138,60){\makebox(0,0)[cc]{$q$}}
%\put(130,28){\makebox(0,0)[cc]{\small ${\log y}$}}
%\put(143,25){\makebox(0,0)[cc]{$p$}}
\put(155,45){\makebox(0,0)[cc]{$\frac 1y$}}
\put(153,35){\makebox(0,0)[cc]{$p$}}
\end{picture}
}
$$
$$
\hbox{
\begin{picture}(0,40)(240,30)
\thicklines
\put(110,40){\makebox(0,0)[cc]{$=$}}
\put(120,40){\makebox(0,0)[cc]{$r$}}
\put(125,40){\circle*{1.5}}
\put(125,40){\vector(1,0){25}}
\put(150,40){\vector(-1,1){15}}
\put(150,40){\line(1,-1){10}}
\put(150,40){\circle*{1.5}}
\put(135,55){\circle*{1.5}}
\put(160,30){\circle*{1.5}}
\put(129,57){\makebox(0,0)[cc]{$\frac{y'}{y}$}}
\put(138,60){\makebox(0,0)[cc]{$q$}}
\put(152,25){\makebox(0,0)[cc]{\small ${\log y}$}}
\put(165,30){\makebox(0,0)[cc]{$p$}}
\put(155,45){\makebox(0,0)[cc]{$\frac 1y$}}
\put(150,35){\makebox(0,0)[cc]{$s$}}
\end{picture}
\begin{picture}(0,40)(160,30)
\thicklines
\put(110,40){\makebox(0,0)[cc]{$-$}}
\put(120,40){\makebox(0,0)[cc]{$r$}}
\put(125,40){\circle*{1.5}}
\put(125,40){\vector(1,0){25}}
\put(150,40){\vector(-1,1){15}}
%\put(150,40){\line(-1,-1){10}}
\put(150,40){\circle*{1.5}}
\put(135,55){\circle*{1.5}}
%\put(140,30){\circle*{1.5}}
\put(129,57){\makebox(0,0)[cc]{$\frac{y'}{y}$}}
\put(138,60){\makebox(0,0)[cc]{$q$}}
%\put(130,28){\makebox(0,0)[cc]{\small ${\log y}$}}
%\put(143,25){\makebox(0,0)[cc]{$p$}}
\put(155,45){\makebox(0,0)[cc]{$\frac {y'}{y^2}$}}
\put(153,35){\makebox(0,0)[cc]{$s$}}
\end{picture}
\begin{picture}(0,40)(100,30)
\thicklines
\put(110,40){\makebox(0,0)[cc]{$+$}}
\put(120,40){\makebox(0,0)[cc]{$r$}}
\put(125,40){\circle*{1.5}}
\put(125,40){\line(1,0){25}}
\put(150,40){\vector(1,1){15}}
%\put(150,40){\line(-1,-1){10}}
\put(150,40){\circle*{1.5}}
\put(165,55){\circle*{1.5}}
%\put(140,30){\circle*{1.5}}
\put(162,57){\makebox(0,0)[cc]{$q$}}
\put(171,60){\makebox(0,0)[cc]{$\frac{y'}{y}$}}
%\put(130,28){\makebox(0,0)[cc]{\small ${\log y}$}}
%\put(143,25){\makebox(0,0)[cc]{$p$}}
\put(150,47){\makebox(0,0)[cc]{$\frac 1y$}}
\put(153,35){\makebox(0,0)[cc]{$p$}}
\end{picture}
\begin{picture}(0,40)(40,30)
\thicklines
\put(110,40){\makebox(0,0)[cc]{$+$}}
\put(120,40){\makebox(0,0)[cc]{$r$}}
\put(125,40){\circle*{1.5}}
\put(125,40){\line(1,0){25}}
%\put(150,40){\vector(-1,1){15}}
%\put(150,40){\line(-1,-1){10}}
\put(150,40){\circle*{1.5}}
%\put(135,55){\circle*{1.5}}
%\put(140,30){\circle*{1.5}}
%\put(129,57){\makebox(0,0)[cc]{$\frac{y'}{y}$}}
%\put(138,60){\makebox(0,0)[cc]{$q$}}
%\put(130,28){\makebox(0,0)[cc]{\small ${\log y}$}}
%\put(143,25){\makebox(0,0)[cc]{$p$}}
\put(155,45){\makebox(0,0)[cc]{$\frac {y'}{y^2}$}}
\put(153,35){\makebox(0,0)[cc]{$s$}}
\end{picture}
}
$$
$$
\hbox{
\begin{picture}(0,40)(240,30)
\thicklines
\put(110,40){\makebox(0,0)[cc]{$=-$}}
\put(120,40){\makebox(0,0)[cc]{$r$}}
\put(125,40){\circle*{1.5}}
\put(125,40){\vector(1,0){25}}
\put(150,40){\vector(1,1){10}}
\put(150,40){\vector(1,-1){10}}
\put(150,40){\circle*{1.5}}
\put(160,50){\circle*{1.5}}
\put(160,30){\circle*{1.5}}
\put(165,50){\makebox(0,0)[cc]{$\frac{y'}{y}$}}
\put(158,55){\makebox(0,0)[cc]{$q$}}
\put(165,30){\makebox(0,0)[cc]{$\frac{y'}{y}$}}
\put(158,25){\makebox(0,0)[cc]{$p$}}
\put(149,47){\makebox(0,0)[cc]{$\frac 1y$}}
\put(149,35){\makebox(0,0)[cc]{$s$}}
\end{picture}
\begin{picture}(0,40)(160,30)
\thicklines
\put(110,40){\makebox(0,0)[cc]{$-2$}}
\put(120,40){\makebox(0,0)[cc]{$r$}}
\put(125,40){\circle*{1.5}}
\put(125,40){\vector(1,0){20}}
%\put(150,40){\vector(-1,1){15}}
%\put(150,40){\line(-1,-1){10}}
\put(145,35){\makebox(0,0)[cc]{$s$}}
\put(145,47){\makebox(0,0)[cc]{$\frac{y'}{y^2}$}}
\put(145,40){\circle*{1.5}}
\put(145,40){\vector(1,0){20}}
\put(165,40){\circle*{1.5}}
\put(170,40){\makebox(0,0)[cc]{$\frac {y'}{y}$}}
\put(165,35){\makebox(0,0)[cc]{$p$}}
\end{picture}
\begin{picture}(0,40)(80,30)
\thicklines
\put(110,40){\makebox(0,0)[cc]{$-$}}
\put(120,40){\makebox(0,0)[cc]{$r$}}
\put(125,40){\circle*{1.5}}
\put(125,40){\vector(1,0){25}}
\put(150,40){\circle*{1.5}}
\put(163,40){\makebox(0,0)[cc]{$\frac {y'^2}{y^3}$}}
\put(150,35){\makebox(0,0)[cc]{$s$}}
\end{picture}
}
$$
$$
\hbox{
\begin{picture}(0,40)(240,30)
\thicklines
\put(110,40){\makebox(0,0)[cc]{$-$}}
\put(120,40){\makebox(0,0)[cc]{$r$}}
\put(125,40){\circle*{1.5}}
\put(125,40){\vector(1,0){20}}
\put(143,35){\makebox(0,0)[cc]{$p$}}
\put(160,40){\makebox(0,0)[cc]{$\d_p\biggl(\frac{1}{y(p)}$}}
\put(175,40){\vector(1,0){20}}
\put(195,40){\circle*{1.5}}
\put(192,45){\makebox(0,0)[cc]{$q$}}
\put(205,40){\makebox(0,0)[cc]{$\frac{y'}{y}\biggr)$}}
\end{picture}
\begin{picture}(0,40)(120,30)
\thicklines
\put(110,40){\makebox(0,0)[cc]{$-$}}
\put(120,40){\makebox(0,0)[cc]{$r$}}
\put(125,40){\circle*{1.5}}
\put(125,40){\vector(1,0){25}}
\put(150,40){\circle*{1.5}}
\put(163,40){\makebox(0,0)[cc]{$\d_s\left(\frac {y'}{y^2}\right)$}}
\put(150,35){\makebox(0,0)[cc]{$s$}}
\end{picture}
}
$$
$$
\hbox{
\begin{picture}(0,40)(240,30)
\thicklines
\put(110,40){\makebox(0,0)[cc]{$=-$}}
\put(120,40){\makebox(0,0)[cc]{$r$}}
\put(125,40){\circle*{1.5}}
\put(125,40){\vector(1,0){25}}
\put(150,40){\vector(1,1){10}}
\put(150,40){\vector(1,-1){10}}
\put(150,40){\circle*{1.5}}
\put(160,50){\circle*{1.5}}
\put(160,30){\circle*{1.5}}
\put(165,50){\makebox(0,0)[cc]{$\frac{y'}{y}$}}
\put(158,55){\makebox(0,0)[cc]{$q$}}
\put(165,30){\makebox(0,0)[cc]{$\frac{y'}{y}$}}
\put(158,25){\makebox(0,0)[cc]{$p$}}
\put(149,47){\makebox(0,0)[cc]{$\frac 1y$}}
\put(149,35){\makebox(0,0)[cc]{$s$}}
\end{picture}
\begin{picture}(0,40)(160,30)
\thicklines
\put(110,40){\makebox(0,0)[cc]{$-$}}
\put(120,40){\makebox(0,0)[cc]{$r$}}
\put(125,40){\circle*{1.5}}
\put(125,40){\vector(1,0){20}}
\put(143,35){\makebox(0,0)[cc]{$p$}}
\put(160,40){\makebox(0,0)[cc]{$\frac{1}{y(p)}\d_p\biggl($}}
\put(175,40){\vector(1,0){20}}
\put(195,40){\circle*{1.5}}
\put(192,45){\makebox(0,0)[cc]{$q$}}
\put(205,40){\makebox(0,0)[cc]{$\frac{y'}{y}\biggr)$}}
\end{picture}
}
$$
$$
\hbox{
\begin{picture}(0,40)(240,30)
\thicklines
\put(110,40){\makebox(0,0)[cc]{$-$}}
\put(120,40){\makebox(0,0)[cc]{$r$}}
\put(125,40){\circle*{1.5}}
\put(125,40){\vector(1,0){20}}
%\put(150,40){\vector(-1,1){15}}
%\put(150,40){\line(-1,-1){10}}
\put(145,35){\makebox(0,0)[cc]{$s$}}
\put(145,47){\makebox(0,0)[cc]{$\frac{y'}{y^2}$}}
\put(145,40){\circle*{1.5}}
\put(145,40){\vector(1,0){20}}
\put(165,40){\circle*{1.5}}
\put(170,40){\makebox(0,0)[cc]{$\frac {y'}{y}$}}
\put(165,35){\makebox(0,0)[cc]{$p$}}
\end{picture}
\begin{picture}(0,40)(160,30)
\thicklines
\put(110,40){\makebox(0,0)[cc]{$-$}}
\put(120,40){\makebox(0,0)[cc]{$r$}}
\put(125,40){\circle*{1.5}}
\put(125,40){\vector(1,0){25}}
\put(150,40){\circle*{1.5}}
\put(170,40){\makebox(0,0)[cc]{$\left(\frac{y''}{y^2}-\frac {y'^2}{y^3}\right)$}}
\put(150,35){\makebox(0,0)[cc]{$s$}}
\end{picture}
}
$$
The first line, together with the second term in (\ref{F02:4}), just give (with the
minus sign) the desired term $W_{0,2}(r)$ (\ref{W02}), the first term in the second line
is canceled by the first term in (\ref{F02:4}), so the only mismatch is due to the
second term in the second line. This term acquires the form
$$
-\oint_{{\cal C}_{\cal D}^{(s)}}dE_{s,\bar s}(r)\frac1{y(s)}\left(\frac{y''(s)}{y(s)}-\frac{y'^2(s)}{y^2(s)}\right)
\equiv-{\widehat{d{\cal E}}}_{r,s}\left(\frac{\d}{\d s}\left(\frac{y'(s)}{y(s)}\right)\right),
$$
where we use the standard notation of \cite{ChEy} (cf. (\ref{Y.2})). But this term can be
already calculated just by taking residues at the branching points, which gives
(for the comprehensive calculation, see \cite{ChMarMirVas2})
\be
\label{F02:5}
\sum_{\alpha=1}^{2n}{\widehat{d{\cal E}}}_{r,s}\left(\frac{1}{2(s-\mu_\alpha)^2}\right)=
-\frac13\frac{\d}{\d V(r)}\log\left(\prod_{\alpha=1}^{2n}M_{\alpha}^{(1)}\Delta(\mu)\right),
\ee
where $M_{\alpha}^{(1)}\equiv \left.M(p)\right|_{p=\mu_\alpha}$ are the first {\em moments}
of the potential and $\Delta(\mu)$ is the Vandermonde determinant in $\mu_\alpha$.

Therefore, combining all the terms, we conclude that
\be
\label{F02:6}
{\cal F}_{0,2}=
-\oint_{{\cal C}_{\cal D}^{(q)}}\frac{dq}{2\pi i}\frac{y'(q)}{y(q)}\int_{D}dE_{q,\bar q}(p)\log y(p) dp
-\frac13\log\left(\prod_{\alpha=1}^{2n}M_{\alpha}^{(1)}\Delta(\mu)\right),
\ee
and we see that, in analogy with the answer obtained by Wigmann and Zabrodin \cite{WZ3}
for the corresponding correction in the normal matrix model case, we have {\em quantum correction}
term similar to the one in ${\cal F}_{1,0}$ (the second term in (\ref{F02:6})).

This completes the calculation of the exceptional term ${\cal F}_{0,2}$ and we therefore have {\em all} terms
of the asymptotic expansion of the eigenvalue model (\ref{X.1}).

\section*{Acknowledgments}

Our work is partly supported by the RFBR grant No.~03-01-22000 (L.Ch.),
by the Grant of Support for the Scientific
Schools NSh-2052.2003.1 (L.Ch.),
and by the Program Mathematical Methods of Nonlinear Dynamics (L.Ch.),
by the Enigma European network MRT-CT-2004-5652 (L.Ch., B.E.), by the ANR project
G\'eom\'etrie et int\'egrabilit\'e en physique math\'ematique ANR-05-BLAN-0029-01 (B.E., L. Ch.),
and by the Enrage European network MRTN-CT-2004-005616 (B.E.).

%%%%%%%%%%%%%%%%%%%%%%%%%%  Bibliography
%%%%%%%%%%%%%%%%%%%%%%%%%%%%%%%%%%


\begin{thebibliography}{99}

\bibitem{Ak96} G.Akemann,
``Higher genus correlators for the Hermitian matrix model with multiple cuts'',
{\em Nucl. Phys.} {\bf B482} (1996) 403, hep-th/9606004

%\bibitem{AkAm} G.Akemann and J.Ambj{\o}rn,
%``New universal spectral correlators'',
%{\em J.Phys.} {\bf A29} (1996) L555--L560, cond-mat/9606129.

\bibitem{ACKM}
J.Ambj{\o}rn, L.Chekhov, C.F.Kristjansen and Yu.Makeenko,
``Matrix model calculations beyond the spherical limit'',
{\em Nucl.Phys.} {\bf B404} (1993) 127--172; Erratum ibid. {\bf B449} (1995) 681,
hep-th/9302014.


\bibitem{loop}
A.A.~Migdal,
%{`Loop equations and $1/N$ expansion'}.
Phys.Rep. {\bf 102} (1983) 199\\
J.~Ambj{\o}rn, J.~Jurkiewicz and Yu.~Makeenko,
%{`Multiloop correlators for two-dimensional quantum gravity'}.
Phys.Lett. {\bf B251} (1990) 517\\
Yu.~Makeenko,
%{\it Loop equations in matrix models and in 2D quantum gravity},
Mod.Phys.Lett. (Brief Reviews) {\bf A6} (1991)
1901--1913

\bibitem{MarcoF} M. Bertola, ''Free Energy of the Two-Matrix
Model/dToda Tau-Function'',
preprint CRM-2921 (2003), hep-th/0306184.


\bibitem{Bertola} M. Bertola,
''Second and Third Order Observables of the Two-Matrix Model'',
{\em J. High Energy Phys.} JHEP11(2003)062 , hep-th/0309192.

%\bibitem{marcosemipot} M. Bertola,
%''Bilinear semi-classical moment functionals and their integral   representation'', math/02051.


%\bibitem{Bertolasem} M. Bertola,
%''Two-matrix model with semiclassical potentials and extended Whitham hierarchy'', hep-th/0511295.

%\bibitem{BDE}  G. Bonnet, F. David, B. Eynard, ``Breakdown of universality in multi-cut matrix models'',
%{\em  J.Phys.} {\bf A33} 6739-6768  (2000).

%\bibitem{BIPZ} E. Brezin, C. Itzykson, G. Parisi, and J. Zuber,
%{\em Comm. Math. Phys.} {\bf 59}, 35  (1978).

\bibitem{Chekh} L.Chekhov,
``Genus one corrections to multi-cut matrix model solutions'',
{\em Theor. Math. Phys. } {\bf 141} (2004) 1640--1653, hep-th/0401089.

%\bibitem{chhard} L.Chekhov,
%''Matrix models with hard walls: Geometry and solutions''; contribution to special volume of J.Phys.A on matrix models, hep-th/0602013.

\bibitem{ChEy}  L.Chekhov and B.Eynard,
''Hermitian matrix model free energy: Feynman graph technique for all genera'',
hep-th/0504116

\bibitem{ChEyOr} L.Chekhov, B.Eynard, and N.Orantin
``Free energy topological expansion for the 2-matrix model,''
hep-th/0603003.

\bibitem{ChMarMirVas}
L.Chekhov, A.Marshakov, A.Mironov, and D.Vasiliev,
''DV and WDVV'',
Phys. Lett. {\bf 562B} (2003) 323--338, hep-th/0301071

\bibitem{ChMarMirVas2}
L.Chekhov, A.Marshakov, A.Mironov, and D.Vasiliev,
''Complex geometry of matrix models,'' {\em Proc. Steklov Inst. Math.}
{\bf 251} (2005) 254--292, hep-th/0506075.

\bibitem{ChMir}
L.Chekhov and A.Mironov,
''Matrix models vs. Seiberg--Witten/Whitham theories'',
Phys.Lett. {\bf 552B} (2003) 293--302, hep-th/0209085

%\bibitem{Virasoro} David F.,
%``Loop equations and nonperturbative effects in two-dimen\-sional quantum gravity''.
%{\em Mod.Phys.Lett.} {\bf A5} (1990) 1019.

\bibitem{Virasoro}
F.David,
{``Loop equations and nonperturbative effects in two-dimen\-sional
quantum gravity''}.
{\em Mod.Phys.Lett.} {\bf A5} (1990) 1019\\
A.Mironov and A.Morozov, Phys.Lett. {\bf B252} (1990) 47-52\\
Ambj{\o}rn J. and Makeenko Yu.,
%{`Properties of loop equations for the Hermitean matrix model and
%for two-dimensional quantum gravity'}.
Mod.Phys.Lett. {\bf A5} (1990) 1753\\
H.Itoyama and Y.Matsuo,
%{\it Noncritical Virasoro algebra of d<1
%matrix model and quantized string field},
Phys.Lett. {\bf 255B} (1991) 202

%\bibitem{dkmvz} P. Deift, T. Kriecherbauer, K. T. R. McLaughlin,
%S. Venakides, Z. Zhou, ``Uniform asymptotics for polynomials
%orthogonal with respect to varying exponential weights and
%applications to universality questions in random matrix theory'',
%{\it Commun. Pure Appl. Math.} {\bf 52}, 1335--1425 (1999).

%\bibitem{ZJDFG} P. Di Francesco, P. Ginsparg, J. Zinn-Justin,
%``2D Gravity and Random Matrices'',
%{\em Phys. Rep.} {\bf 254}, 1 (1995).

\bibitem{DST}
R.Dijkgraaf, A.Sinkovics and M.Tem\"urhan,
{\it Matrix models and gravitational corrections},
{\em Adv.Theor.Math.Phys.} {\bf 7} (2004) 1155--1176;
hep-th/0211241

\bibitem{DV}
R.Dijkgraaf and C.Vafa,
``Matrix Models, Topological Strings, and Supersymmetric Gauge Theories'',
{\em Nucl.Phys.} {\bf 644} (2002) 3--20, hep-th/0206255;
``On Geometry and Matrix Models'',
{\em Nucl.Phys.} {\bf 644} (2002) 21--39, hep-th/0207106;
``A Perturbative Window into Non-Perturbative Physics'',
hep-th/0208048.

\bibitem{Dyson}
F. Dyson, ''Statistical theory of the energy levels of complex systems.'' Parts I, II, and III.
{\em J. Math. Phys.} {\bf 3} (1962) 140, 157, 166.

%\bibitem{DW} R.Dijkgraaf and E.Witten,
%``Mean field theory, topological field theory, and multimatrix models'',
%{\em Nucl.Phys.} {\bf B342} (1990) 486--522.

\bibitem{eynloop1mat} B. Eynard, ``Topological expansion for the 1-hermitian matrix model correlation functions'',
JHEP/024A/0904, xxx, hep-th/0407261.

\bibitem{eynm2m} B. Eynard, ``Large N expansion of the 2-matrix model'',
{\em JHEP} {\bf 01} (2003) 051, hep-th/0210047.

\bibitem{eynm2mg1} B. Eynard, ``Large N expansion of the 2-matrix model,
multicut case'',
preprint SPHT03/106, ccsd-00000521, math-ph/0307052.

\bibitem{EKK}
B.~Eynard, A.~Kokotov, and D.~Korotkin,
``$1/N^2$ corrections to free energy in Hermitian two-matrix model'',
hep-th/0401166.

%\bibitem{eynchain} B. Eynard, ``Eigenvalue distribution of large
%random matrices, from one matrix to several coupled matrices''
%{\em Nucl. Phys. B}{\bf 506}, 633 (1997), cond-mat/9707005.

%\bibitem{eynchaint} B. Eynard, ``Correlation functions of
%eigenvalues of multi-matrix models, and the limit of a time dependent matrix'',
%{\em J. Phys. A: Math. Gen.} {\bf 31}, 8081 (1998), cond-mat/9801075.

%\bibitem{eynmultimat} B. Eynard, ``Master loop equations, free energy and correlations for the chain of matrices'',
% {\em J. High Energy Phys.} JHEP11(2003)018, hep-th/0309036, ccsd-00000572.

%\bibitem{courseynard} B. Eynard ``An introduction to random
%matrices'', lectures given at Saclay, October 2000, notes available at
%http://www-spht.cea.fr/articles/t01/014/.

%\bibitem{courseynardhouches} B. Eynard ``Large N asymptotics of orthogonal polynomials, from integrability to algebraic geometry'',
%Proceeding Les Houches, Applications of Random Matrices in Physics, June 6-25 2004,
%SPhT-T05/036, ccsd-00004536, xxx, math-ph/0503052.

%\bibitem{eynhabilit} B. Eynard, ``Polyn\^omes biorthogonaux, probl\`eme de Riemann-Hilbert et g\'eom\'etrie\\
% alg\'ebrique'', Habilitation \`a diriger les recherches,
%universit\'e Paris VII, (2005).

\bibitem{eylooprat} B.Eynard,
 ``Loop equations for the semiclassical 2-matrix model with hard edges'',
math-ph/0504002.

\bibitem{EOtrmixte} B. Eynard, N. Orantin, ``Mixed correlation functions in the 2-matrix model, and the Bethe ansatz'',
{\em J. High Energy Phys.} JHEP08(2005)028, hep-th/0504029.

\bibitem{EyOran}
B.Eynard, N.Orantin,
''Topological expansion of the 2-matrix model correlation functions: diagrammatic rules for a residue formula'',
 {\em J. High Energy Phys.} JHEP12(2005)034, math-ph/0504058.

\bibitem{Farkas} H.M. Farkas, I. Kra, ''Riemann surfaces'' 2nd edition, Springer Verlag, 1992.

\bibitem{Fay} J.D. Fay, ''Theta functions on Riemann surfaces'', Springer Verlag, 1973.

%\bibitem{Kazakov} V.A. Kazakov, ``Ising model on a dynamical
%planar random lattice: exact solution'',
%{\em Phys Lett.} {\bf A119}, 140-144 (1986).

\bibitem{KazMar} V.A. Kazakov, A. Marshakov, ''Complex Curve of the
Two Matrix Model and its Tau-function'',
{\em J.Phys.} {\bf A36} (2003) 3107-3136, hep-th/0211236.

\bibitem{Kos} I.K.Kostov,
`` Conformal field theory techniques in random matrix models'',
hep-th/9907060.

\bibitem{Kri} I.Krichever
``The $\tau$-function of the universal Whitham
hierarchy, matrix models and topological field theories'',
{\em Commun.Pure Appl.Math.} {\bf47} (1992) 437; hep-th/9205110

\bibitem{WZ2} I. Krichever, M. Mineev-Weinstein, P. Wiegmann, A. Zabrodin,
''Laplacian Growth and Whitham Equations of Soliton Theory'',
nlin.SI/0311005.



%\bibitem{matytsin} A. Matytsin, ``on the large $N$ limit of the
%Itzykson Zuber Integral'',
%{\em Nuc. Phys.} {\bf B411}, 805 (1994), hep-th/9306077.

%\bibitem{Mehta} M.L. Mehta, {\em Random Matrices},2nd edition,
%(Academic Press, New York, 1991).

%\bibitem{staudacher} M. Staudacher,
%`` Combinatorial solution of the 2-matrix model'',
%{\em Phys. Lett.} {\bf B305} (1993) 332-338.

%\bibitem{thooft} G. 't Hooft, {\em Nuc. Phys.} {\bf B72}, 461 (1974).

\bibitem{WZ1} P. Wiegmann, A. Zabrodin,
'Large N expansion for normal and complex matrix ensembles', hep-th/0309253.

\bibitem{WZ3} A. Zabrodin, P. Wiegmann,
''Large N expansion for the 2D Dyson gas'', hep-th/0601009.





\end{thebibliography}
\end{document}